\documentclass[%
reprint,
superscriptaddress,
showpacs,preprintnumbers,
verbatim,
amsmath,amssymb,
aps,
pre,
]{revtex4-1}

\usepackage{color}
\usepackage[pdftex]{graphicx} % Include figure files (pdflatex)
\usepackage{dcolumn}% Align table columns on decimal point
\usepackage{bm}% bold math
\usepackage{textcomp} % to use \textmu

\newcommand{\fsize}{7.1cm}

\newcommand{\dbra}[1]{\left\langle\hspace{-.1cm}\left\langle #1 \right\rangle\hspace{-.1cm}\right\rangle_\nu}
\begin {document}
%section {title}
%\preprint{APS/123-QED}

\title{%
  Generalized Langevin equation with fluctuating diffusivity}
% Force line breaks with \\

%% Time-averaged mean-square-displacement tensor: a novel method to elucidate
%% fluctuating diffusivity

\author{Tomoshige Miyaguchi}
\email{tmiyaguchi@naruto-u.ac.jp}
\affiliation{%
  Department of Mathematics, Naruto University of Education,
  Naruto, Tokushima 772-8502, Japan}

%% \author{Takashi Uneyama}
%% \affiliation{%
%%   % \email{uneyama@mp.pse.nagoya-u.ac.jp}
%% Center for Computational Science, Graduate School of Engineering,
%% Nagoya University, Furo-cho, Chikusa, Nagoya 464-8603,
%% Japan
%% }%

%%   \author{Takuma Akimoto}
%% %   \email{takuma@rs.tus.ac.jp}
%%   \affiliation{%
%%   Department of Physics, Tokyo University of Science, Noda, Chiba 278-8510, Japan
%% }%

%
% for meatadata of arXiv
%
%Brownian motion with alternately fluctuating diffusivity: Stretched-exponential and power-law relaxation
%Tomoshige Miyaguchi, Takashi Uneyama, and Takuma Akimoto
%Phys. Rev. E 100, 012116 (2019)
%10.1103/PhysRevE.100.012116


\date{\today}% It is always \today, today,

%section {abstract}

\begin{abstract}
  A generalized Langevin equation with fluctuating diffusivity (GLEFD) is
  proposed, and it is shown that the GLEFD satisfies a generalized
  fluctuation-dissipation relation. If the memory kernel is a power law, the
  GLEFD exhibits subdiffusion, non-Gaussianity, and
  stretched-exponential relaxation. The case in which the memory kernel is given
  by a single exponential function is also investigated as an analytically
  tractable example. In particular, the mean-square displacement and the
  self-intermediate-scattering function of this system show plateau
  structures. A numerical scheme to integrate the GLEFD is also presented.
\end{abstract}

%\pacs{05.45.Ac, 05.40.Fb, 87.15.Vv}% PACS, the Physics and Astronomy
\maketitle

\section {Introduction}
%subsection {general introduction}
%subsubsection {establish importance of this research topic}

A microscopic tagged-particle suspended in a fluid exhibits random motion called
Brownian motion, which is caused by incessant collisions with solvent molecules
of the fluid
\cite{dhont96}. %Such a microscopic particle undergoing Brownian motion is
%reffered to as a tagged particle.
%
In a dilute and homogeneous condition, the particle shows normal diffusion,
which is characterized by a linear increase of the mean-square displacement
(MSD) with a constant diffusion coefficient.
However, the tagged-particle diffusion in crowded and/or heterogeneous
environments (e.g., cytoplasm of living cells) shows intrinsic statistical
properties such as anomalous diffusion, non-Gaussianity, and random diffusivity
\cite{he08, wang12, parry14, manzo15, lampo17, sabri20, janczura21}. Because cell
functions depend on intracellular transport of macromolecules, elucidation of
these statistical properties is crucially important to understand complex
mechanisms of living cells \cite{berg81, bressloff13}.

%subsubsection {provide general background information}

Anomalous diffusion is characterized by a non-linear increase of the
MSD{\color{black}; in particular, if the MSD increases sublinearly, it is referred
  to as subdiffusion.}  There are several possible mechanisms for such
non-linear behavior. One of the important mechanisms of subdiffusion is a memory
effect or viscoelastic properties of a surrounding medium \cite{goychuk09,
  goychuk12, goychuk21}.  The memory effect originates from interactions with
surrounding molecules, and analytically obtained by a projection operator method
from microscopic equations of motion \cite{kubo91, gotze_book,
  hansen90}. Dynamics of a tagged particle with the memory effect is described
by the generalized Langevin equation (GLE); the memory effect is represented by
an integral term with a memory kernel.  If the memory kernel is a power-law
form, the MSD shows subdiffusion, whereas the displacement is always Gaussian,
because the GLE is a Gaussian process (if there is no external force).

In contrast, the random diffusivity is usually generated either by dynamics of
internal modes of the tagged particle or by heterogeneity of the medium.  For
example, center-of-mass motions of several polymer models show the fluctuating
diffusivity due to dynamics of internal modes \cite{uneyama15, miyaguchi17,
  nampoothiri21, yamamoto21, nampoothiri22}.  As an example of the
heterogeneity, diffusion on random potentials also gives rise to the random
diffusivity \cite{miyaguchi11b, miyaguchi15a}. One of important consequences of
the random diffusivity is the non-Gaussianity of the particle
displacement. However, the MSD for the random diffusivity models shows only
normal diffusion, if the system is in equilibrium \cite{miyaguchi16, jain16,
  chechkin17, tyagi17, lanoislee18}.

%% A single particle, reffered to as a tagged particle, shows slower diffusion
%% compared with the diffusion at infinite dilution. This is caused by the cage
%% formed with particles surrounding the tagged one. In fact, the mean square
%% displacement (MSD) of the tagged particle shows non-linear increase.  Such
%% nonlinear increase of the MSD might be the fingerprint of memory effect, which
%% can be described by generalized Langevin equation or fractional Brownian motion.
%% % 
%% Another important feature of the tagged particle motion is the non-Gaussianity
%% of the displacement and fluctuating diffusivity. One of the typical mechanisms
%% of fluctuating diffusivity is the dynamics of internal degrees of freedom
%% \cite{miyaguchi17}.

%subsection {literature review to solve the general problem}
%subsubsection {describe general problem or current research focus of the field}
Single-particle-tracking experiments \cite{he08, parry14, manzo15, sabri20,
  janczura21} and molecular dynamics simulations \cite{akimoto11, yamamoto14,
  jeon16, yamamoto17} show that tagged particles exhibit subdiffusion,
fluctuating diffusivity, and non-Gaussianity. Thus it is important to
incorporate the fluctuating diffusivity into diffusion processes with a memory
effect such as the GLE.
%subsubsection {provide a brief overview of key research projects}
In fact, to describe the diffusion process observed in complex media such as the
cytoplasm, phenomenological models which combine these two effects have recently
been proposed and studied \cite{chechkin17, slezak18, wang20, wang20b, sabri20,
  janczura21, dieball22, goswami22}.
%subsubsection {describe a gap in the research}
However, physical backgrounds of these phenomenological models have been still
unclear.  For example, the fluctuation-dissipation relation and the detailed
balance should be satisfied if the system is in equilibrium \cite{kou08}, but it
has not been elucidated so far whether these properties hold for the
phenomenological models.

%% It is thus important to elucidate the interactions between particles with
%% fluctuating diffusivity, and .
%% To elucidate the origin of such statistical properties, other than microscopic
%% approaches such as the mode coupling theory, some phenomenological models has
%% been proposed and studied.

%subsection {describe the present paper}
%subsubsection {describe the present paper}
This article proposes and investigates a GLE with fluctuating diffusivity
(GLEFD).
%subsubsection {describe the methodology reported in the present paper}
By using a Markovian embedding method \cite{goychuk09, goychuk12}, the fluctuating
diffusivity is incorporated into an overdamped GLE.
%subsubsection {announce the findings}
Importantly, it is shown that the GLEFD satisfies a generalized
fluctuation-dissipation relation. Moreover, a numerical scheme for the GLEFD is
also presented on the basis of the Markovian embedding.

Then, as an important example, the GLEFD with a power-law memory kernel is
investigated, and show that this system exhibits both subdiffusion and
non-Gaussianity. In addition, as one of the simplest models, we study the GLEFD
with an exponential memory kernel, which is referred to as a dimer model. Due to
its simplicity, this dimer model can be analytically tractable to some extent.
In fact, analytical formulas for the MSD, a non-Gaussian parameter (NGP), and a
self-intermediate-scattering function (ISF) are derived. Furthermore, as a
special class of the dimer model, we also study the case in which the diffusion
coefficient switches between two values, $D_+$ and $D_-$, alternately, and the
distributions of the sojourn times in these two states are given by an
exponential distribution or a power law. {\color{black}In these special classes of
  the GLEFD, it is assumed that a short-time diffusivity is not fluctuating, and
  thus only a long-time diffusivity is fluctuating.}

%subsubsection {organization of the paper}

This paper is organized as follows. In Sec.~\ref{s.gle}, the GLE and the Markovian
embedding are briefly summarized; then, we incorporate the fluctuating
diffusivity into the GLE, and thereby defining the GLEFD. The numerical scheme
of the GLEFD is also presented in this section. In Sec.~\ref{s.fbm}, the case in
which the memory kernel is a power-law form is studied. Moreover,
Secs.\ref{s.dimer} and \ref{s.two-state-model} are devoted to investigation of
the dimer model, of which the memory kernel is a single exponential function.
Finally, in Sec.~\ref{s.discussion}, a discussion on the significance and
implication of the results is presented.  Some technical matters including
numerical setups are presented in Appendices.

\section {Generalized Langevin equation with fluctuating diffusivity}\label{s.gle}
%subsection {intro}

The Langevin equation with fluctuating diffusivity has been studied intensively
during the last decade \cite{mykyta14, uneyama15, miyaguchi17}, and its
properties are now very well known. The memory kernel of this model is given by
a delta function and thus there is no integral term in this equation, whereas a
difficulty in introducing the fluctuating diffusivity into the GLE is in the
presence of {\color{black}an} integral term. This motivates us to transform the
GLE into a form without the integral term by the Markovian embedding method, and
then to introduce the fluctuating diffusivity into this Markovian system.

\subsection {Generalized Langevin equation}

First, let us briefly introduce an overdamped GLE (without fluctuating
diffusivity) of the following form:
\begin{equation}
  \label{e.gle.wo.fd}
  \zeta\frac {d \bm{r}(t)}{dt} + \zeta \int_0^t \phi(t-t') \dot{\bm{r}}(t')dt'
  =
  {\color{black}\bm{F}_{e}} + \bm{\xi}(t) + {\color{black}\bm{\xi}_c^0(t)},
\end{equation}
where the tagged-particle position $\bm{r}(t)$ and an external force
$\bm{F}_{e}(\bm{r}, t) = -\partial V(\bm{r},t)/\partial \bm{r}$ are $n$
dimensional vectors \footnote{ At least in this article, the dimensionarity
  plays no important role. However, the dimensionarity becomes important when
  the diffusivity is represented by a rank two tensor; the tensorial diffusivity
  is necessary to study polymer dynamics \cite{doi86, uneyama15, miyaguchi17}
  and colloid dynamics with hydrodynamic interactions \cite{dhont96,
    miyaguchi20a}.}, and the overdot in $\dot{\bm{r}}$ denotes the time
derivative. {\color{black}Moreover, $\bm{\xi}(t)$ and $\bm{\xi}_c^0(t)$ in
  Eq.~(\ref{e.gle.wo.fd}) are $n$-dimensional Gaussian noises, and if the system
  is in equilibrium, they should satisfy the fluctuation-dissipation relations:
\begin{align}
  \label{e.<xi-xi>}
  \left\langle \bm{\xi}(t) \bm{\xi}(t') \right\rangle
  &=
  2k_BT \zeta \delta (t-t') \bm{I},
  \\[0.1cm]
  \label{e.<xi_c-xi_c>}
  \left\langle \bm{\xi}_c^0(t) \bm{\xi}_c^0(t') \right\rangle
  &=
  k_BT \zeta \phi (t-t') \bm{I}.
\end{align}
}
Here $k_B$ is the Boltzmann constant, $T$ is the absolute temperature, and
$\bm{I}$ is the $n\times n$ identity tensor. Moreover, $\delta(t)$ is the delta
function, and $\phi(t)$ is the memory kernel, which is an even function of
$t$. Accordingly, $\bm{\xi}(t)$ is a white noise, and $\bm{\xi}_c^0(t)$ a
correlated noise; {\color{black}the superscript $0$ in $\bm{\xi}_c^0(t)$ indicates
  the absence of the fluctuating diffusivity.} $\zeta$ is the friction constant
with respect to the white noise, but $\zeta$ is also put in front of the
integral term in Eq.~(\ref{e.gle.wo.fd}) just to make $\phi(t)dt$ dimensionless.

In some literature, the model without the friction term $\zeta \dot{\bm{r}}(t)$
or the white Gaussian noise $\bm{\xi}(t)$ in Eq.~(\ref{e.gle.wo.fd}) is studied
\cite{deng09, jeon12b, kou08}. For the monomer dynamics of the Rouse model,
however, it is shown that these terms naturally arise \cite{panja10b}. In
addition, if the friction term $\zeta \dot{\bm{r}}(t)$ and the white Gaussian
noise $\bm{\xi}(t)$ in Eq.~(\ref{e.gle.wo.fd}) are absent, the integral equation
(\ref{e.gle.wo.fd}) does not have a solution $\bm{v}(t) = \dot{\bm{r}}(t)$ in
general \cite{doetsch12book}.
%% ; for example, if $\phi(t)$ is power-law form $t^{-\alpha}$ with
%% $0<\alpha<1$, $\bm{v}(t)$ cannot be obtained.
%% [a solution $\bm{v}(t)$ exists if there is an external field $\bm{F}_e $
%% \cite{kou08} or if $\phi(0) < \infty$ \cite{doetsch12book}; for the latter case,
%% an equation similar to Eq.~(\ref{e.gle.wo.fd}) can be obtained by
%% differentiating the equation with $t$].
Thus, the incorporation of these terms might well be physically natural. In
addition, the friction term $\zeta \dot{\bm{r}}(t)$ plays an essential role in
constructing Markovian equations of motion below.

For simplicity, we rewrite Eq.~(\ref{e.gle.wo.fd}) as
\begin{equation}
  \label{e.gle.wo.fd.simple}
  \frac {d \bm{r}}{dt} + \int_0^t \phi(t-t') \dot{\bm{r}}(t')dt'
  =
  \beta D\bm{F}_{e} + \sqrt{2D}\bm{\xi} + \sqrt{D}\bm{\xi}_c^0,
\end{equation}
where $D := k_BT/\zeta$ is the diffusion coefficient with respect to the white
Gaussian noise $\bm{\xi}(t)$, and $\beta = 1/k_BT$ is the inverse temperature.
In Eq.~(\ref{e.gle.wo.fd.simple}), $\bm{\xi}(t)$ and $\bm{\xi}_c^0(t)$ are
redefined as $\bm{\xi} \to \bm{\xi} (2 k_BT \zeta)^{1/2}$ and
{\color{black}$\bm{\xi}_c^0 \to \bm{\xi}_c^0 (k_BT \zeta)^{1/2}$} instead of
defining new variables.  Accordingly, the fluctuation-dissipation relations
[Eqs.~(\ref{e.<xi-xi>}) and (\ref{e.<xi_c-xi_c>})] are modified as
\begin{align}
  \label{e.<xi-xi>.simple}
  \left\langle \bm{\xi}(t) \bm{\xi}(t') \right\rangle
  &=
  \delta (t-t') \bm{I},
  \\[0.1cm]
  \label{e.<xi_c-xi_c>.simple}
  \left\langle \bm{\xi}_c^0(t) \bm{\xi}_c^0(t') \right\rangle
  &=
  \phi (t-t') \bm{I}.
\end{align}

\subsection {Markovian embedding}\label{s.markov-embed}

To incorporate fluctuating diffusivity into Eq.~(\ref{e.gle.wo.fd.simple}), we
use the Markovian embedding method \cite{goychuk09}. First, the correlated noise
$\bm{\xi}_c^0(t)$ is approximated as a superposition of stochastic processes
$\bm{\xi}_i(t)$ as
\begin{equation}
  \label{e.def.xi_c^0}
  \bm{\xi}_c^0(t) \approx \sum_{i=0}^{N-1} \bm{\xi}_i(t),
\end{equation}
where $\bm{\xi}_i(t)$ is assumed to satisfy the equation for the
Ornstein-Uhlenbeck process:
\begin{equation}
  \label{e.dxi_i(t)/dt}
  \frac {d \bm{\xi}_i(t) }{dt}
  =
  - \nu_i \bm{\xi}_i(t) + \sqrt{2k_i' \nu_i } \bm{\eta}_i(t).
\end{equation}
Here, $\nu_i$ and $k_i'$ are positive constants to be specified below
[Eq.~(\ref{e.phi(t)_approx})], and $\bm{\eta}_i(t)$ is a white Gaussian noise with
\begin{equation}
  \label{e.<eta_i(t)eta_j(t')>}
  \left\langle \bm{\eta}_i(t)\bm{\eta}_j(t') \right\rangle
  =
  \delta_{ij} \delta(t-t') \bm{I}.
\end{equation}
We also assume that $\bm{\eta}_i(t)$ and $\bm{\xi}(t)$ are independent, and that
$\bm{\xi}_i(t)$ is in equilibrium, and thus $\bm{\xi}_i(t)$ follows a canonical
distribution $\propto\exp(- \bm{\xi}_i^2/2k'_i)$; the latter assumption is
necessary to make $\bm{\xi}_c^0(t)$ a stationary process
[Eq.~(\ref{e.<xi_c-xi_c>.simple})].

Then, from Eqs.~(\ref{e.dxi_i(t)/dt}) and (\ref{e.<eta_i(t)eta_j(t')>}), a
correlation matrix for $\bm{\xi}_i(t)$ is given by
\begin{equation}
  \label{e.<xi_i-xi_j>}
  \left\langle \bm{\xi}_i(t)\bm{\xi}_j(t') \right\rangle
  =
  k_i' \delta_{ij} e^{-\nu_i|t-t'|} \bm{I}.
\end{equation}
From Eqs.~(\ref{e.<xi_c-xi_c>.simple}), (\ref{e.def.xi_c^0}), and
(\ref{e.<xi_i-xi_j>}), we have an approximation to the correlation function
$\phi(t)$ as
\begin{equation}
  \label{e.phi(t)_approx}
  \phi(t) \approx \sum_{i=0}^{N-1} k_i' e^{-\nu_i |t|}.
\end{equation}
The parameters $k_i'$ and $\nu_i$ are determined so that
Eq.~(\ref{e.phi(t)_approx}) holds approximately. Thus, the above equation is an
approximation to $\phi(t)$ with a superposition of exponential functions, but if
$N$ tends to infinity, the approximation is expected to become exact
\cite{goychuk09}.

Next, to deal with the integral term in Eq.~(\ref{e.gle.wo.fd.simple}), let us
define auxiliary variables $\bm{r}_i(t)$ as \cite{goychuk12}
\begin{equation}
  \label{e.(ri-r)}
  k_i'(\bm{r}_i - \bm{r})
  =
  \sqrt{D}\bm{\xi}_i - k_i' \int_0^t e^{-\nu_i (t-t')} \dot{\bm{r}}(t') dt',
\end{equation}
Inserting Eqs.~(\ref{e.def.xi_c^0}) and (\ref{e.phi(t)_approx}) into
Eq.~(\ref{e.gle.wo.fd.simple}), we have the Markovian equations of motion as
\begin{align}
  \label{e.gle.wo.fd.markov.1}
  \frac {d \bm{r}(t)}{dt}
  &=
  \beta D\bm{F}_{e} + \sqrt{2D}\bm{\xi}(t)
  -\beta D\sum_{i=0}^{N-1} k_i(\bm{r}-\bm{r}_i),\\[0.1cm]
  \label{e.gle.wo.fd.markov.2}
  \frac {d \bm{r}_i(t)}{dt} 
  &=
  - \beta D_ik_i (\bm{r}_i- \bm{r}) + \sqrt{2D_i} \bm{\eta}_i(t), 
\end{align}
where the second equation is obtained by differentiating Eq.~(\ref{e.(ri-r)})
with time $t$.  Here, $k_i:= k_i'/\beta D$ corresponds to a spring constant, and
$D_i:= \nu_i/k_i\beta$ is a diffusivity of the auxiliary variable $\bm{r}_i$.
In this way, the GLE [Eq.~(\ref{e.gle.wo.fd})] is transformed into a Markovian
system [Eqs.~(\ref{e.gle.wo.fd.markov.1}) and (\ref{e.gle.wo.fd.markov.2})], in
which the tagged particle $\bm{r}$ interacts with the auxiliary variables
$\bm{r}_i$ through harmonic potentials. In addition,
Eqs.~(\ref{e.<xi-xi>.simple}) and (\ref{e.<eta_i(t)eta_j(t')>}) are the
fluctuation-dissipation relations for Eqs.~(\ref{e.gle.wo.fd.markov.1}) and
(\ref{e.gle.wo.fd.markov.2}).  If the friction term $d \bm{r}(t)/dt$ is absent
in Eq.~(\ref{e.gle.wo.fd.simple}), then Eq.~(\ref{e.gle.wo.fd.markov.1}) becomes
an algebraic equation.

The Markovian embedding of the underdamped GLE is presented in
Ref.~\cite{goychuk09}, whereas Eqs.~(\ref{e.gle.wo.fd.markov.1}) and
(\ref{e.gle.wo.fd.markov.2}) are a Markovian embedding for the overdamped GLE.
Obviously, Eqs.~(\ref{e.gle.wo.fd.markov.1}) and (\ref{e.gle.wo.fd.markov.2})
can be readily utilized as a numerical scheme to simulate the non-Markovian
equation (\ref{e.gle.wo.fd.simple}). It is also the case even if we introduce
the fluctuating diffusivity into Eq.~(\ref{e.gle.wo.fd.markov.2}) as shown
below.

{\color{black}In this work, any physical meanings of the auxiliary variables are
  not considered explicitly, but they should be related to fast variables which
  are projected out from microscopic equations of motion. This is because the
  GLE is usually obtained by projecting out fast variables, and $\phi(t)$
  and $\bm{\xi}_c^0(t)$ in the GLE [Eq.~(\ref{e.gle.wo.fd.simple})] are related
  to these fast variables \cite{hansen90}. As an example, it is known that the
  middle monomer of the Rouse polymer chain can be described by
  Eqs.~(\ref{e.gle.wo.fd.markov.1}) and (\ref{e.gle.wo.fd.markov.2}) with
  suitable choices of $D_i$ and $k_i$ \cite{panja10b}. In this example, the
  auxiliary variables are normal modes of the Rouse polymer, and the kernel
  $\phi(t)$ is a power-law function with an exponential cutoff.}

\subsection {Fluctuating diffusivity}

Now, we incorporate the fluctuating diffusivity into
Eq.~(\ref{e.gle.wo.fd.markov.2}) as
\begin{equation}
  \label{e.gle.w.fd.markov.1}
  \frac {d \bm{r}_i(t)}{dt} 
  =
  - \beta k_i D_i(t) (\bm{r}_i- \bm{r}) + \sqrt{2D_i(t)} \bm{\eta}_i(t), 
\end{equation}
where the fluctuating diffusivity $D_i(t)$ is a stochastic process. We assume
that $D_i(t)$ and $\bm{\eta}_j(t)$ are independent for any $i$ and
$j$. {\color{black}In contrast, Eq.~(\ref{e.gle.wo.fd.markov.1}) remains
  unchanged. Note that the fluctuation-dissipation relation and the detailed
  balance hold for Eqs.~(\ref{e.gle.wo.fd.markov.1}) and
  (\ref{e.gle.w.fd.markov.1}), and thus, if $D_i(t)$ are stationary processes as
  well, the whole process is in equilibrium.

  Next, let us derive an integro-differential equation similar to
  Eq.~(\ref{e.gle.wo.fd.simple}).} According to the modification from
Eq.~(\ref{e.gle.wo.fd.markov.2}) to Eq.~(\ref{e.gle.w.fd.markov.1}),
Eqs.~(\ref{e.dxi_i(t)/dt}) and (\ref{e.<xi_i-xi_j>}) are also modified
respectively as
\begin{align}
  \label{e.dxi_i(t)/dt.w.fd}
  \frac {d \bm{\xi}_i(t) }{dt}
  &=
  - \nu_i(t) \bm{\xi}_i(t) + \sqrt{2k_i'\nu_i(t)} \bm{\eta}_i(t),
  \\[0.1cm]
  \label{e.<xi_i-xi_j>.w.fd}
  \left\langle \bm{\xi}_i(t)\bm{\xi}_j(t') \right\rangle
  &=
  k_i' \delta_{ij} \bm{I} e^{- \left|\int_{t'}^{t}\nu_i(u)du\right|},
\end{align}
where $\nu_i(t) := \beta k_iD_i(t)$ and $\bm{\eta}_i(t)$ is the white Gaussian
noise satisfying Eq.~(\ref{e.<eta_i(t)eta_j(t')>}). Thus, $\nu_i(t)$ is the
inverse of an fluctuating relaxation time, but, for simplicity, $\nu_i(t)$ is
also referred to as the fluctuating diffusivity.

From Eqs.~(\ref{e.gle.w.fd.markov.1}) and (\ref{e.dxi_i(t)/dt.w.fd}), we obtain
\begin{equation}
  \label{e.gle.w.fd.markov.2}
  k_i'\frac {d}{dt}\left(\bm{r}_i - \bm{r}\right) 
  =
  - k_i'\dot{\bm{r}}
  - k_i'\nu_i(t) (\bm{r}_i- \bm{r})
  + \sqrt{D} \left[ \dot{\bm{\xi}}_i+ \nu_i(t) \bm{\xi}_i \right].  
\end{equation}
A solution of this equation is given by
\begin{equation}
  \label{e.gle.w.fd.markov.3}
  k_i'\left(\bm{r}_i - \bm{r}\right) 
  =
  \sqrt{D}{\bm{\xi}}_i
  -
  k_i'\int_0^t e^{-\int_{t'}^{t} \nu_i(u) du} \dot{\bm{r}}(t')dt', 
\end{equation}
as can be checked directly by differentiating Eq.~(\ref{e.gle.w.fd.markov.3}) in
terms of $t$. Inserting Eq.~(\ref{e.gle.w.fd.markov.3}) into
Eq.~(\ref{e.gle.wo.fd.markov.1}), we finally obtain the GLEFD
\begin{equation}
  \label{e.glefd}
  \frac {d \bm{r}}{dt}
  + 
  \int_0^t \phi(t,t') \dot{\bm{r}}(t')dt' 
  =
  \beta D\bm{F}_{e} + \sqrt{2D}\bm{\xi}
  + \sqrt{D}\bm{\xi}_c,
\end{equation}
with a generalized fluctuation-dissipation relation
\begin{equation}
  \label{e.phi(t,t')}
  \left\langle \bm{\xi}_c(t)\bm{\xi}_c(t') \right\rangle
  = \phi(t,t') \bm{I} 
  := \bm{I} \sum_{i=0}^{N-1} k_i'e^{-\left|\int_{t'}^{t} \nu_i(u) du \right|}. 
\end{equation}
The memory kernel $\phi(t,t')$, which is defined by the second equality in
Eq.~(\ref{e.phi(t,t')}), depends on the fluctuating diffusivity $\nu_i(u)$.  In
other words, $\phi(t,t')$ is a functional of $\nu_i(t)$. {\color{black} As in the
  case of $\bm{\xi}_c^0(t)$ [Eq.~(\ref{e.def.xi_c^0})], $\bm{\xi}_c(t)$ in
  Eq.~(\ref{e.glefd}) is given by
\begin{equation}
  \label{e.def.xi_c}
  \bm{\xi}_c(t) = \sum_{i=0}^{N-1} \bm{\xi}_i(t),
\end{equation}
where $\bm{\xi}_i(t)$ is redefined by Eqs.~(\ref{e.dxi_i(t)/dt.w.fd}) and
(\ref{e.<xi_i-xi_j>.w.fd}). Therefore, the diffusivity due to $\bm{\xi}_c(t)$ is
fluctuating; the absence of the superscript $0$ in $\bm{\xi}_c(t)$ indicates the
presence of the fluctuating diffusivity. } Moreover,
Eqs.~(\ref{e.gle.wo.fd.markov.1}) and (\ref{e.gle.w.fd.markov.1}) can be
utilized as a numerical scheme for the GLEFD [Eq.~(\ref{e.glefd})],

{\color{black}In Eq.~(\ref{e.phi(t,t')}), it looks as if time-translation
  invariance is violated, but it is not the case. This apparent violation is due
  to the fact that Eq.~(\ref{e.phi(t,t')}) is a formula for a single realization
  of diffusivity paths [$\nu_i(t)$]. If the diffusivities $\nu_i(t)$ are
  stationary processes and initially in equilibrium, then, by taking ensemble
  averages over the diffusivity paths and the initial conditions in
  Eq.~(\ref{e.phi(t,t')}), we have a fluctuation-dissipation relation with the
  time-translation invariance as
  \begin{equation}
    \label{e.<phi(t,t')>_nu}
    \left\langle
    \left\langle \bm{\xi}_c(t)\bm{\xi}_c(t') \right\rangle
    \right\rangle_{\nu}
    = \phi_{\nu}(t-t') \bm{I}
    := \bm{I} \sum_{i=0}^{N-1} k_i'
    \left\langle
    e^{-\left|\int_{t'}^{t} \nu_i(u) du \right|}
    \right\rangle_{\nu},
  \end{equation}
  where $\phi_{\nu}(t-t')$ is defined by the right-hand side of the above
  equation. }

%% \begin{align}
%%   \label{e.glefd}
%%   \frac {d \bm{r}(t)}{dt}
%%   =&
%%   \beta D\bm{F}_{e} + \sqrt{2D}\bm{\xi}
%%   + \sqrt{D}\sum_{i=1}^N\bm{\xi}_i
%%   \notag\\[0.0cm]
%%   &- 
%%   \int_0^t \sum_{i=1}^N k_i'e^{-\int_{t'}^{t} \nu_i(s) ds} \dot{\bm{r}(t')}dt'. 
%% \end{align}

{\color{black}Note also that the diffusivity $D$ with respect to the white noise
  $\bm{\xi}(t)$ remains constant in Eq.~(\ref{e.glefd}), because a goal of this
  work was to incorporate the fluctuating diffusivity into the correlated noise
  $\bm{\xi}_c^0(t)$ in Eq.~(\ref{e.gle.wo.fd.simple}). Moreover, the} correlated
noise $\bm{\xi}_c(t)$ in Eq.~(\ref{e.glefd}) is defined through the
Ornstein-Uhlenbeck processes $\bm{\xi}_i(t)$ with fluctuating diffusivity
[Eq.~(\ref{e.dxi_i(t)/dt.w.fd})], but the variance of $\bm{\xi}_i(t)$ and hence
also the variance of $\bm{\xi}_c(t)$ are independent of time due to the detailed
balance. Thus, only their relaxation behavior is fluctuating, but, as shown in
the next section, the long-time (generalized) diffusivity of $\bm{r}(t)$ is
fluctuating; this is why we refer to the model defined by Eq.~(\ref{e.glefd}) as
the GLEFD.

{\color{black}Due to coexistence of trajectories with different diffusivity paths
  $D_i(u)$, the position vectors $\bm{r}_i(t)$ and $\bm{r}(t)$ are non-Gaussian
  processes \cite{uneyama19}. For fixed paths $D_i(u)$ [or, equivalently, fixed
  $\nu_i(u)$], however, $\bm{r}_i(t)$ and $\bm{r}(t)$ are Gaussian processes (if
  the external force $\bm{F}_e$ is absent). This is because
  Eqs.~(\ref{e.gle.wo.fd.markov.1}) and (\ref{e.gle.w.fd.markov.1}) are linear
  in $\bm{r}_i(t)$ and $\bm{r}(t)$.} Thanks to this pathwise Gaussianity, the
ISF for the dimer model can be easily derived as shown in Sec.~\ref{s.dimer}. If
we take into account more than {\color{black}one realization} of the diffusivity
paths $[D_i(u)]$, $\bm{r}_i(t)$ and $\bm{r}(t)$ become non-Gaussian processes.

In Ref.~\cite{slezak18}, the GLE with a random parameter is proposed, but the
random parameter is assumed to be time independent. Namely, the model in
Ref.~\cite{slezak18} is a compound GLE, in which GLEs with different parameter
values are simply superimposed. In Eq.~(\ref{e.glefd}), however, the random
diffusivity $\nu_i(t)$ is time dependent and fluctuating, and this makes the
theoretical analysis of Eq.~(\ref{e.glefd}) quite difficult. In fact, the
integral term in Eq.~(\ref{e.glefd}) is not a convolution, and thus it is
impossible to use the Laplace transformation, which is a standard method to
analyze the GLE \cite{pottier03, burov08}.

\section {Fractional Brownian motion with fluctuating diffusivity}\label{s.fbm}
%subsection {intro}

In this section, we study the GLEFD with a power-law memory kernel, which might
well be important to explain subdiffusion and non-Gaussianity observed in
biological experiments \cite{sabri20, janczura21}. Numerical simulations show
that the GLEFD with a power-law memory kernel exhibits subdiffusion,
non-Gaussianity, and stretched-exponential relaxation.

\subsection {Fractional Brownian motion}

First, let us study the GLE [Eq.~(\ref{e.gle.wo.fd.simple})] without the
fluctuating diffusivity or the external force $\bm{F}_e$:
\begin{equation}
  \label{e.gle.wo.Fe}
  \frac {d \bm{r}}{dt}
  + 
  \int_0^t \phi(t-t') \dot{\bm{r}}(t')dt' 
  =
  \sqrt{2D}\bm{\xi} + \sqrt{D}\bm{\xi}_c^0.
\end{equation}
Here, the memory kernel $\phi(t-t')$ is assumed to be given by a power law
\begin{equation}
  \label{e.phi(t).power}
  \phi(t)
  =
  \frac {D}{D_{\alpha}}
  \frac {t^{-\alpha}}{\Gamma(1+\alpha) \Gamma(1-\alpha)} = A t^{-\alpha},
\end{equation}
where $\alpha$ is the power-law index with $0 < \alpha < 1$, $\Gamma(\dots)$ is
the Gamma function, {\color{black}$D_{\alpha}$ is a generalized diffusion
  coefficient of subdiffusion}, and
$A := D/D_{\alpha}\Gamma(1+\alpha) \Gamma(1-\alpha)$.

{\color{black}It is often important to take into account cutoffs of the power-law
  kernel in studying real phenomena \cite{pottier03, panja10b, molina18}. In
  addition, in approximating the power-law kernel with
  Eq.~(\ref{e.phi(t)_approx}), short- and long-time cutoffs are unavoidable as
  explained shortly. Therefore, the numerical scheme proposed in the previous
  section entails errors caused by these cutoffs at short and long times [but,
  the short-time cutoff might not matter due to the presence of the white noise
  $\bm{\xi}(t)$ in Eq.~(\ref{e.gle.wo.fd.simple})]. Here, however, we focus only
  on pure power-law time regimes, in which cutoff effects are negligible
  \cite{goychuk09}.}

With the memory kernel defined in Eq.~(\ref{e.phi(t).power}), the GLE without
the fluctuating diffusivity [Eq.~(\ref{e.gle.wo.Fe})] shows normal diffusion at
short time and subdiffusion at long time as \cite{panja10b, pottier03} (See also
Appendix \ref{app.fbm})
\begin{align}
  \label{e.gle.msd}
  \left\langle \delta\bm{r}^2(t) \right\rangle
  \simeq
  \begin{cases}
    2n D t                   & t\ll t_c, \\
    2n D_{\alpha} t^{\alpha} & t\gg t_c,
  \end{cases}
\end{align}
where $\delta \bm{r}(t)$ is a displacement vector defined as
$\delta \bm{r}(t):= \bm{r}(t) - \bm{r}(0)$, and a crossover time $t_c$ is
$t_c:= [\Gamma(1+\alpha)D_\alpha/D]^{1/(1-\alpha)}$. More precisely, the
displacement correlation $\left\langle \bm{r}(t)\bm{r}(t') \right\rangle$ is
given by
$\left\langle \bm{r}(t)\bm{r}(t') \right\rangle \simeq D_{\alpha}(t^{\alpha} +
t'^{\alpha} - |t-t'|^{\alpha})$ at long times, and thus hereafter this model is
referred to as the fractional Brownian motion (FBM), although short-time
behavior is different from the original FBM \cite{kou08}.

%subsection {markov enbedding}
To obtain the Markovian embedding of Eq.~(\ref{e.gle.wo.Fe}), let us put
$k_i' = A' \nu_i^{\alpha}$ with $A':= A (b-1) / b^{1/2}\Gamma(\alpha) $ in
Eq.~(\ref{e.phi(t)_approx}), and then the power-law kernel
[Eq.~(\ref{e.phi(t).power})] is approximated as
\begin{equation}
  \label{e.phi(t).power.markov}
  \phi(t) \approx A' \sum_{i=0}^{N-1}\nu_i^{\alpha} e^{-\nu_i|t|},
\end{equation}
where $\nu_i := \nu_0/b^i$ with a constant $b>1$ and a high frequency cutoff
$\nu_0$ \cite{goychuk09}. Then, Eq.~(\ref{e.phi(t).power.markov}) can be
approximated by an integral as
$\approx A'\int_{0}^{\infty} \nu^{\alpha-1} e^{-\nu t}d\nu b^{1/2} /(b-1) =
A't^{-\alpha}\Gamma(\alpha) b^{1/2} /(b-1) = At^{-\alpha}$, and therefore we
obtain Eq.~(\ref{e.phi(t).power}). This integral approximation is exact at
$b \to 1$ and $N \to \infty$, but, it is found that the choices $b=2$, $N=128$,
and $\nu_0 = 1000 / \mu_+$ are sufficiently good to describe
Eq.~(\ref{e.phi(t).power}) in numerical simulations for the two-state model
defined below [Eq.~(\ref{e.glefd.two-state})]. See Appendix \ref{app:nondim} for
simulation detail.

\subsection {Fluctuating diffusivity: two-state model}

Here, we derive and analyze the GLEFD [Eq.~(\ref{e.glefd})] with a power-law
memory kernel; this model is referred to as FBM with fluctuating diffusivity
(FBMFD). {\color{black}The equation of motion is given by Eq.~(\ref{e.glefd})
  without the external force $\bm{F}_e$
\begin{equation}
  \label{e.glefd.wo.Fe}
  \frac {d \bm{r}}{dt}
  + 
  \int_0^t \phi(t,t') \dot{\bm{r}}(t')dt' 
  =
  \sqrt{2D}\bm{\xi}(t)
  + \sqrt{D}\bm{\xi}_c(t).
\end{equation}
Because of the correspondence between Eq.~(\ref{e.phi(t)_approx}) and
Eq.~(\ref{e.phi(t,t')}), the memory kernel $\phi(t,t')$ is obtained by modifying
Eq.~(\ref{e.phi(t).power.markov}) as
\begin{equation}
  \label{e.phi(t,t').power}
  \phi(t,t') :=
  A' \sum_{i=0}^{N-1} \nu_i^{\alpha}
  e^{-\left|\int_{t'}^{t} \nu_i(u) du\right|}.
\end{equation}}

As a simple case study, let us focus on the FBMFD with a two-state
diffusivity. The two-state diffusivity has been used frequently to explain
statistical properties of single-particle-tracking data \cite{sabri20,
  janczura21, dieball22}.  The two-state diffusivity is defined by
\begin{equation}
  \label{e.glefd.two-state}
  \nu_i(t) = \nu_i \kappa (t) = 
  \begin{cases}
    \nu_i \kappa_+, \\[0.1cm]
    \nu_i \kappa_-,
  \end{cases}
\end{equation}
where $\nu_i(t) = \beta k_iD_i(t)$ as before, $\nu_i$ are constants with the
same dimension as $\nu_i(t)$, and $\kappa(t)$ is a dimensionless two-state
process \cite{godrche01, miyaguchi19, miyaguchi16}.

More precisely, the diffusivity $\nu_i(t)$ is assumed to switch between the two
states at random times $t_1, t_2,\dots$. Let
$\tau_k := t_k - t_{k-1} \,(k=1,2,\dots)$ be sojourn times of the two states,
where we define $t_0 = 0$ for convenience.
The $+$ ($-$) state is referred to as the fast (slow) state, and the sojourn
time distribution of the fast (slow) state is denoted as $\rho^+(\tau)$
[$\rho^-(\tau)$]. In the numerical simulations of the FBMFD, $\rho^{\pm} (\tau)$
are assumed to follow exponential distributions with means $\mu_{\pm}$
[Eq.~(\ref{e.exp-dist})].

By using Eq.~(\ref{e.glefd.two-state}), Eq.~(\ref{e.phi(t,t')}) is rewritten as
%% \begin{equation}
%%   \label{e.phi(t,t').two-state}
%%   \phi(t,t') := \sum_{i=0}^{N-1} k_i'e^{-\nu_i\int_{t'}^{t} \sigma(s) ds}. 
%% \end{equation}
%% For the GLEFD with the two-state diffusivity [Eq.~(\ref{e.phi(t,t').two-state})],
%% we set
\begin{equation}
  \label{e.phi(t,t').two-state.power}
  \phi(t,t') :=
  A' \sum_{i=0}^{N-1} \nu_i^{\alpha}
  e^{-\nu_i\left|\int_{t'}^{t} \kappa(u) du\right|}, 
\end{equation}
where $k_i'$ is set as $k_i' = A' \nu_i^{\alpha}$ as in the case of the FBM. If
there is no switching, the diffusive state is $\nu_i(t) \equiv \nu_i\kappa_+$ or
$\nu_i(t) \equiv \nu_i\kappa_-$ for all $t$ depending on the initial state. It
follows that, if the initial state is $\pm$, the MSD grows as
$\left\langle \delta\bm{r}^2(t) \right\rangle \simeq 2n D_{\alpha}^{\pm}
t^{\alpha}$ with $D^{\pm}_{\alpha} = D_{\alpha} \kappa_{\pm}^{\alpha}$ at long
time. If the diffusive state switches very slowly between the two states, the
diffusivity might be regarded as being fluctuating between the two diffusion
coefficients $D_{\alpha}^{\pm}$.

It is not that simple in general, however, because, if the diffusive state
switches from one state to the other, information of the previous state is not
lost immediately due to the memory effect represented by the integral in
Eq.~(\ref{e.phi(t,t').two-state.power}).  In fact, by setting
$\nu_i = \nu_0/b^i$ and using the integral approximation mentioned above,
Eq.~(\ref{e.phi(t,t').two-state.power}) can be expressed as
\begin{equation}
  \label{e.phi(t,t').two-state.power.approx}
  \phi(t,t')
  \approx
  \frac {D}{D_{\alpha}}
  \frac
  {\left|\int_{t'}^t \kappa(u)du\right|^{-\alpha}}
  {\Gamma(1+\alpha) \Gamma(1-\alpha)}.
\end{equation}
Note that, if $\kappa(t)$ is independent of time $t$,
Eq.~(\ref{e.phi(t,t').two-state.power.approx}) represents an algebraic decay
$\sim |t-t'|^{-\alpha}$, thus it is a generalization of
Eq.~(\ref{e.phi(t).power}).

%subsection {fig1}
\begin{figure}[t!]
  \centerline{\includegraphics[width=\fsize]{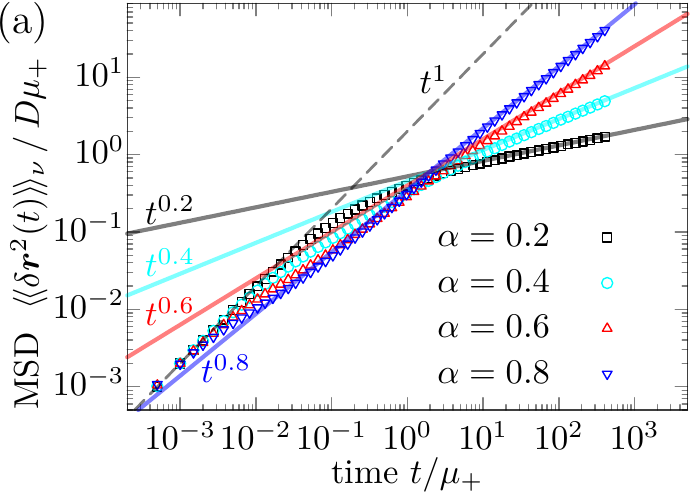}}
  \centerline{\includegraphics[width=\fsize]{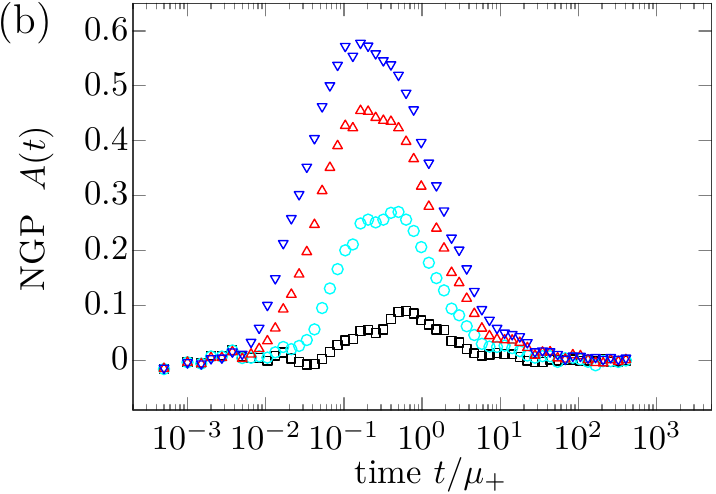}}
  \centerline{\includegraphics[width=\fsize]{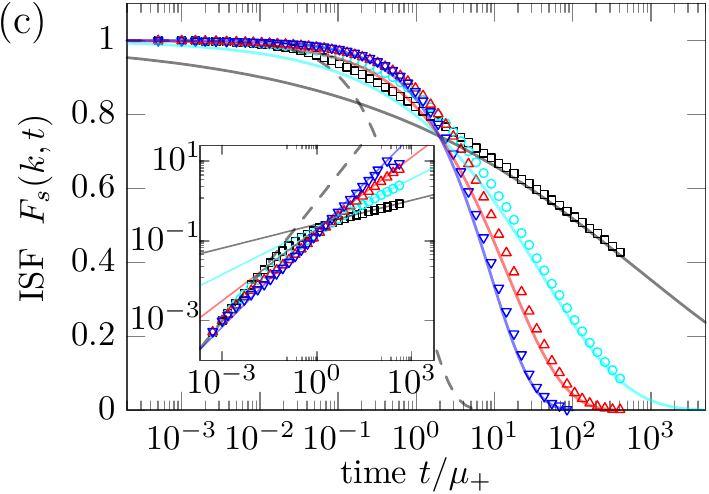}}
  \caption{\label{f.fbm.alp} FBMFD for four different values of $\alpha$:
    $\alpha = 0.2$, $0.4$, $0.6$, and $0.8$. The symbols are data obtained by numerical
    integration of Eqs.~(\ref{e.gle.wo.fd.markov.1}) and
    (\ref{e.gle.w.fd.markov.1}) with the correlation function in
    Eq.~(\ref{e.phi(t,t').two-state.power}). The generalized diffusivity
    $D_\alpha$ is fixed as $D_\alpha/D\mu_+^{1-\alpha} = 0.3$, the fast and slow
    state diffusivities $\kappa_\pm$ as $(\kappa_+, \kappa_-) = (1.0, 0.01)$,
    and the mean sojourn time $\mu_-$ as $\mu_- = \mu_+$.
    (a) MSD vs time. The dashed line and the full lines are the
    short-time and long-time predictions in Eq.~(\ref{e.fbmfd.msd}).  (b) NGP vs
    time. (c) ISF vs time. The dashed line and the full lines are the short-time
    and long-time predictions in Eq.~(\ref{e.fbmfd.isf}). The wave number $k$ is
    set as $k = 1/\sqrt{D\mu_+}$. {\color{black} (inset) $\log[F_s(k,t)]$ vs $t$
      in log-log form.}  [In (b-c), results for the four different values of
    $\alpha$ are displayed with the same color code as in (a)].  }
\end{figure}

%subsection {fig2}
\begin{figure}[t!]
  \centerline{\includegraphics[width=\fsize]{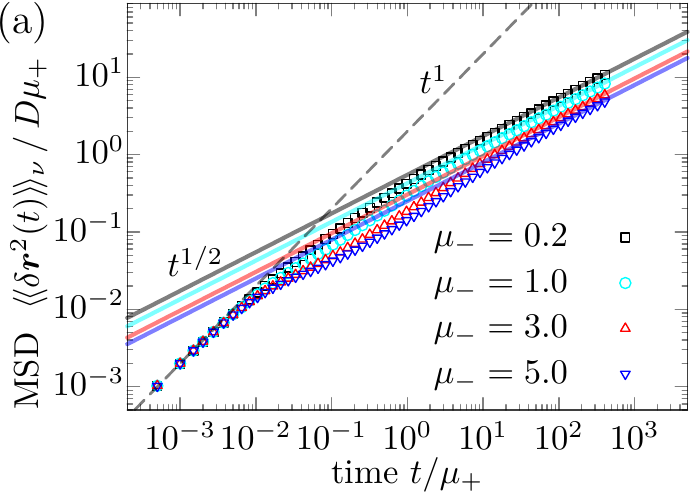}}
  \centerline{\includegraphics[width=\fsize]{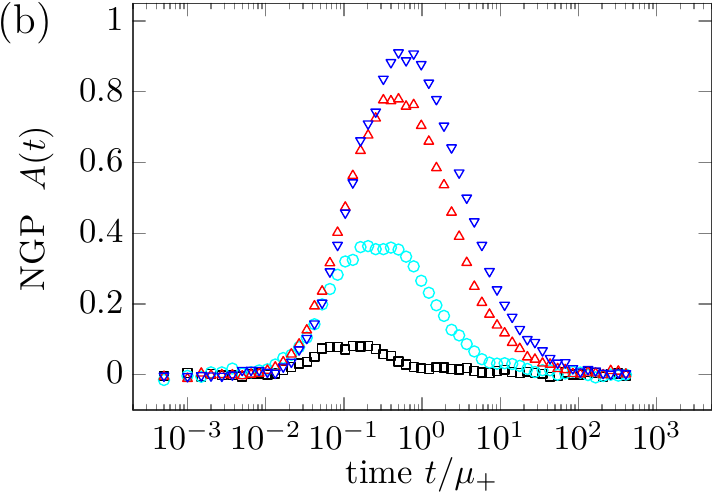}}
  \centerline{\includegraphics[width=\fsize]{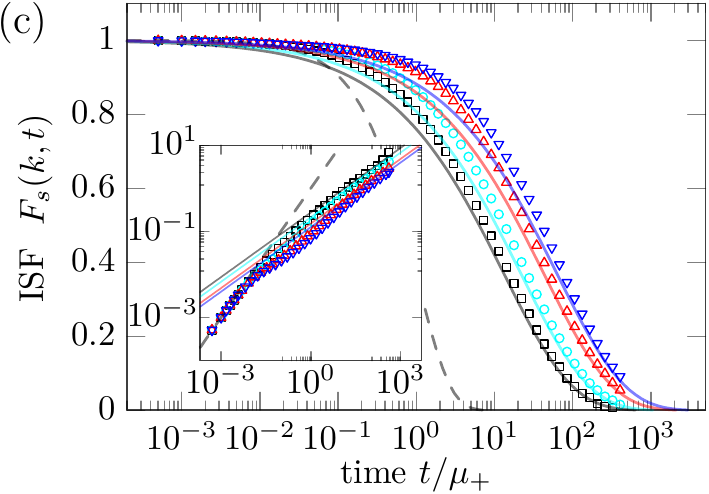}}
  \caption{\label{f.fbm.ms} FBMFD for four different values of $\mu_-$:
    $\mu_-/\mu_+ = 0.2$, $1.0$, $3.0$, and $5.0$. The power-law index $\alpha$
    is fixed as $\alpha = 0.5$. Parameters $D_\alpha$, $\kappa_\pm$ and $k$ are
    the same as those in Fig.~\ref{f.fbm.alp}.  (a) MSD vs time. The dashed line
    and the full lines are the short-time and long-time predictions in
    Eq.~(\ref{e.fbmfd.msd}). (b) NGP vs time. (c) ISF vs time. The dashed line
    and the full lines are the short-time and long-time predictions in
    Eq.~(\ref{e.fbmfd.isf}).  {\color{black} (inset) $\log[F_s(k,t)]$ vs $t$ in
      log-log form.} [In (b-c), results for the four different values of $\mu_-$
    are displayed with the same color code as in (a)].  }
\end{figure}

%subsection {results}
If $t-t'$ is much larger than the mean sojourn times $\mu_{\pm}$, and
$\kappa(t)$ is a stationary process with mean
$\left\langle \kappa \right\rangle$, the integral in
Eq.~(\ref{e.phi(t,t').two-state.power.approx}) can be approximated as
$\int_{t'}^t \kappa(s) ds \approx \left\langle \kappa \right\rangle (t-t')$.
Thus, for large $t-t'$, we obtain 
\begin{equation}
  \label{e.phi(t,t').two-state.power.approx.long-time}
  \phi(t,t')
  \approx
  \frac {D}{D_{\alpha}^{\mathrm{eq}}}
  \frac
  {|t-t'|^{-\alpha}}
  {\Gamma(1+\alpha) \Gamma(1-\alpha)},
\end{equation}
where
$D_{\alpha}^{\mathrm{eq}} := D_{\alpha}\left\langle \kappa
\right\rangle^{\alpha}$. Because this memory kernel is the same form as that in
Eq.~(\ref{e.phi(t).power}), it follows that the long-time behavior is also the
same as that of Eq.~(\ref{e.gle.msd}) with the diffusivity
$D_{\alpha}^{\mathrm{eq}}$. Moreover, the memory kernel does not affect the
short-time dynamics, and thus the short-time behavior is exactly the same as
that in Eq.~(\ref{e.gle.msd}).

It follows that we have
\begin{align}
  \label{e.fbmfd.msd}
  \left\langle \delta\bm{r}^2(t) \right\rangle
  \simeq
  \begin{cases}
    2n D t                                 & t\ll t_c', \\[.15cm]
    2n D_{\alpha}^{\mathrm{eq}} t^{\alpha} & t\gg t_c',
  \end{cases}
\end{align}
where $t_c'$ is defined by
$t_c' := [\Gamma(1+\alpha)D_{\alpha}^{\mathrm{eq}}/D]^{1/(1-\alpha)}$ [See
Appendix \ref{app.fbm}].  Therefore, the MSD show subdiffusion at long time
limit $t \to \infty$; this is contrasting to the model studied in
Refs.~\cite{wang20, wang20b}, for which a crossover to normal diffusion at long
time is observed.

{\color{black}From Eq.~(\ref{e.fbmfd.msd}), it can be seen that the short-time
  diffusivity is not fluctuating, whereas the long-time diffusivity should be
  fluctuating, because it contains a mean value
  $\left\langle \kappa \right\rangle$. Moreover, there} is no qualitative
difference in the MSD between the systems with and without the fluctuating
diffusivity. As a result, to obtain the information on the fluctuating
diffusivity for the present model, higher order moments such as the NGP should
be investigated. This property that no information on the fluctuating
diffusivity can be obtained by the second order moment is a remarkable feature
of the {\color{black}Langevin equation with fluctuating diffusivity}, in which the
memory kernel is the delta function \cite{uneyama15, miyaguchi17}. This
prominent property remains to be valid also for the GLEFD.

Finally, let us consider the ISF $F_s(k, t)$, which is a Fourier transform of
the displacement distribution $G(\delta\bm{r}, t)$ at time $t$; namely,
$F_s(k, t) := \int_{-\infty}^{\infty}d\delta\bm{r}\, e^{-i \bm{k}\cdot
  \delta\bm{r}} G(\delta\bm{r}, t)$ with $k := | \bm{k}|$. {\color{black}Note that
  the ISF is referred to as a characteristic function in probability theory.}
For short and long time regimes ($t\ll t_c'$ and $t\gg t_c'$), the process might
well be approximated as a Gaussian process, and therefore the ISF is simply
given by
\begin{align}
  \label{e.fbmfd.isf}
  F_s(k, t)
  \simeq
  \begin{cases}
    \exp(-k^2 D t)                                 & t\ll t_c', \\[.15cm]
    \exp(-k^2 D_{\alpha}^{\mathrm{eq}} t^{\alpha}) & t\gg t_c',
  \end{cases}
\end{align}
where the formulas for the MSD in Eq.~(\ref{e.fbmfd.msd}) are used. Thus, the
ISF exhibits a stretched-exponential relaxation at long time.

\subsection {Numerical results}

In Fig.~\ref{f.fbm.alp}, numerical results for the MSD $\dbra{\delta \bm{r}^2}$,
the NGP $A(t)$, and the ISF $F_s(k, t)$ are displayed for different values of
$\alpha$. Here, the NGP $A(t)$ is defined by
\begin{equation}
  \label{e.ngp-def}
  A(t)
  :=
  \frac {n}{n+2}
  \frac
  {\dbra{\delta \bm{r}^4}}
  {\dbra{\delta \bm{r}^2}^2} - 1,
\end{equation}
{\color{black}with $n$ being the space dimension.}  In numerical simulations, we
assume that the sojourn times $\tau$ for the fast and slow states follow the
exponential distributions with means $\mu_{\pm}$. In Fig.~\ref{f.fbm.ms},
numerical results for different values of $\mu_-$ are presented.

For the MSD, the numerical results displayed in Figs.~\ref{f.fbm.alp}(a) and
\ref{f.fbm.ms}(a) are consistent with the theoretical prediction in
Eq.~(\ref{e.fbmfd.msd}). A difference between the FBM (without the fluctuating
diffusivity) and the FBMFD is prominent in non-Gaussianity; for the FBM, the NGP
is zero because it is a Gaussian process, whereas for the FBMFD, the NGP is
positive at an intermediate time scale around which the transition from normal
diffusion to subdiffusion occurs. As shown in Figs.~\ref{f.fbm.alp}(b) and
\ref{f.fbm.ms}(b), the NGP $A(t)$ shows a unimodal shape, and $A(t)$ is larger
for $\alpha$ closer to unity and larger values of $\mu_-$ (the latter tendency
is easy to understand, because $\mu_-=0$ corresponds to a system without the
fluctuating diffusivity). As displayed in Fig.~\ref{f.fbm.alp}(c) and
\ref{f.fbm.ms}(c), the ISF shows exponential relaxation at short time (the
dashed line) and stretched-exponential relaxation at long time (the full
lines). Numerical results depicted by symbols are consistent with the prediction
Eq.~(\ref{e.fbmfd.isf}).  {\color{black}Moreover, in the insets of
  Fig.~\ref{f.fbm.alp}(c) and \ref{f.fbm.ms}(c), $\log[-F_s(k,t)]$ is displayed
  in log-log form; a close resemblance with the MSD [Fig.~\ref{f.fbm.alp}(a) and
  \ref{f.fbm.ms}(a)] can be clearly seen. This resemblance is due to the fact
  that the process is Gaussian at small and large $t$ as shown in
  Figs.~\ref{f.fbm.alp}(b) and \ref{f.fbm.ms}(b).}

\section {Dimer model}\label{s.dimer}
%subsection {intro}

In general, it is difficult to analytically study the GLEFD
[Eq.~(\ref{e.glefd})], because its integral term is not a convolution. Thus,
here we focus on the simplest case in which there is only a single relaxation
mode (i.e., $N=1$). This system can be analyzed with elementally manipulations,
and, in fact, analytical expressions for the MSD, the NGP, and the ISF are
derived under some assumptions. These analytical results would give some
insights for the complex interplay between the memory effect and the fluctuating
diffusivity.

{\color{black} The equation of motion without the fluctuating diffusivity is given
  by Eq.~(\ref{e.gle.wo.Fe}), and the memory kernel $\phi(t)$ is assumed to have
  a single relaxation mode
  \begin{equation}
    \label{e.dimer.phi(t)}
    \phi(t)
    =
    k'_0e^{-\nu_0 \left|t\right|}. 
  \end{equation}
  The associated GLEFD is given by Eq.~(\ref{e.glefd.wo.Fe}), and, from the
  correspondence between Eq.~(\ref{e.phi(t)_approx}) and
  Eq.~(\ref{e.phi(t,t')}), $\phi(t,t')$ is obtained by modifying
  Eq.~(\ref{e.dimer.phi(t)}) as
  \begin{equation}
    \label{e.dimer.phi(t,t')}
    \phi(t,t')
    =
    \left\langle \bm{\xi}_c(t)\bm{\xi}_c(t') \right\rangle
    =
    k'_0e^{-\left|\int_{t'}^{t} \nu_0(u) du\right|}. 
  \end{equation}}

With the Markovian embedding method [Eqs.~(\ref{e.gle.wo.fd.markov.1}) and
(\ref{e.gle.w.fd.markov.1})], Eq.~(\ref{e.glefd.wo.Fe}) can be rewritten as
\begin{align}
  \label{e.dimer.dr(t)/dt}
  \frac {d \bm{r}(t)}{dt}
  &=
  -\beta D k_0 (\bm{r} - \bm{r}_0) + \sqrt{2D} \bm{\xi}(t),
  \\[0.1cm]
  \label{e.dimer.dr1(t)/dt}
  \frac {d \bm{r}_0(t)}{dt}
  &=
  -\beta D_0(t) k_0 (\bm{r}_0 - \bm{r}) + \sqrt{2D_0(t)} \bm{\eta}_0(t),
\end{align}
where $\nu_0(t):= \beta k_0 D_0(t)$ and $k_0:= k_0' / \beta D$ as before. Thus,
this is a model with two particles interacting through a harmonic potential, and
thus it is referred to as a dimer model in the following. {\color{black}
  Short-time dynamics of $\bm{r}(t)$ is obviously given by
  $\dot{\bm{r}} \sim \sqrt{2D} \bm{\xi}$, and, if $D_0(t) \ll D$, its long-time
  dynamics is described by $\dot{\bm{r}} \sim \sqrt{2D_0(t)} \bm{\eta}_0$. Thus,
  the short-time diffusivity is not fluctuating, but the long-time diffusivity
  is fluctuating.}

A model similar to Eqs.~(\ref{e.dimer.dr(t)/dt}) and (\ref{e.dimer.dr1(t)/dt})
was studied in Ref.~\cite{straeten11}, but it does not satisfy the
fluctuation-dissipation relation. Hence, the model in Ref.~\cite{straeten11} is
intrinsically out of equilibrium; in contrast, Eqs~(\ref{e.dimer.dr(t)/dt}) and
(\ref{e.dimer.dr1(t)/dt}) are in equilibrium (if the initial ensemble is in
equilibrium).
It is also interesting that similar models are studied as a toy model of
supercooled liquids and glassy systems \cite{Fodor16, hachiya19}. In such
models, $\bm{r}(t)$ and $\bm{r}_0(t)$ might represent position coordinates of a
tagged particle and of a cage which confines the tagged particle,
respectively. A cage is formed by particles surrounding the tagged one, and
hence it might be physically natural to assume that the diffusion of the cage
position is much slower than that of the tagged particle, i.e., $D_0(t) \ll
D$. In the following, we derive some formulas for the dimer model under this
assumption.
%% In the models studied in these papers, the non-Gaussian noise \cite{Fodor16}
%% and random trappings \cite{hachiya19} are utilized instead of the fluctuating
%% diffusivity.

\subsection {Representation without memory kernel}\label{s.represent-xi_all}

%% $\left\langle ... \right\rangle$ stands for the ensemble average over the noises
%% [$\bm{\xi}_i(t)$] and/or initial conditions. 

%% To eliminate $\bm{r}_2(t)$ from Eq.~(\ref{e.dimer.dr(t)/dt}), Eq.~(\ref{e.dimer.dr1(t)/dt})
%% is solved formally, and we obtain
%% \begin{align}
%%   \Delta \bm{r}(t)
%%   %   \bm{r}_2(t) - \bm{r}_1(t)
%%   &=
%%   %   [\bm{r}_2(0) - \bm{r}_1(0)] e^{- u\int_0^t dt_1 D_2(t_1)}
%%   \Delta \bm{r}(0) e^{- u\int_0^t dt_1 D_2(t_1)}
%%   - \int_0^t dt_2 \dot{\bm{r}}(t_2) e^{-u\int_0^t dt_1 D_2(t_1)}
%%   \notag\\[0.1cm]
%%   &+ \int_0^t dt_2 \sqrt{2 D_2(t_2)} \bm{\xi}_2 (t_2)
%%   e^{-u\int_t^{t_2} dt_1 D_2(t_1)},
%% \end{align}
%% where $\Delta \bm{r}(t)$ is defined as
%% $\Delta \bm{r}(t):= \bm{r}_2(t) - \bm{r}_1(t)$.

Here, we derive yet another expression of the equation of motion, which is
useful for later analysis. Subtraction of Eq.~(\ref{e.dimer.dr1(t)/dt}) from
Eq.~(\ref{e.dimer.dr(t)/dt}) gives
\begin{equation}
  \label{e.clefd_cage}
  \frac {d }{dt} \Delta \bm{r}(t)
  =
  - \beta k_0 D_{\mathrm{ou}}(t)  \Delta \bm{r}(t)
  + \sqrt{2D_{\mathrm{ou}}(t) } \bm{\xi}_{\mathrm{ou}}(t),
\end{equation}
where $\Delta \bm{r}(t)$ is the difference of the two position vectors
$\Delta \bm{r}(t):=\bm{r}(t) - \bm{r}_0(t)$, and the diffusivity
$D_{\mathrm{ou}}(t)$ is defined as $D_{\mathrm{ou}}(t) := D +
D_0(t)$. Accordingly, the noise terms are rewritten as
$\sqrt{D}\bm{\xi}(t) - \sqrt{D_0(t)}\bm{\eta}(t) = \sqrt{D_{\mathrm{ou}}(t)}
\bm{\xi}_{\mathrm{ou}}(t)$. It follows that $\bm{\xi}_{\mathrm{ou}}(t)$ is also
a white Gaussian noise and satisfies
\begin{align}
  \left\langle \bm{\xi}_{\mathrm{ou}}(t) \bm{\xi}_{\mathrm{ou}}(t')  \right\rangle
  &= \delta (t-t') \bm{I},
  \\[0.1cm]
  \label{e.<xi(t)xi_ou(t')>}
  \left\langle \bm{\xi}(t) \bm{\xi}_{\mathrm{ou}}(t') \right\rangle
  &= \sqrt{\frac {D}{D_{\mathrm{ou}}(t)}}\delta (t-t')
  \bm{I}.
\end{align}
The equation (\ref{e.clefd_cage}) is the Ornstein-Uhlenbeck process with the
fluctuating-diffusivity; {\color{black}this model is studied by Fox \cite{fox77}
  for deterministic $D_{\mathrm{ou}}(t)$, and it is intensively studied in
  Refs.~\cite{uneyama19, miyaguchi19} for stochastic $D_{\mathrm{ou}}(t)$.}

The solution of Eq.~(\ref{e.clefd_cage}) is expressed as
\begin{align}
  \Delta\bm{r}(t)
  &=
  \Delta\bm{r}_0 e^{-\int_{0}^{t} \nu_{\mathrm{ou}}(u) du}
  \notag\\[0.1cm]
  \label{e.r(t)-R(t).clefd}
  &+
  \int_{0}^{t} dt_1
  \sqrt{2 D_{\mathrm{ou}}(t_1)} \bm{\xi}_{\mathrm{ou}}(t_1)
  e^{-\int_{t_1}^{t} \nu_{\mathrm{ou}}(u)du},
\end{align}
where $\Delta\bm{r}_0 := \Delta\bm{r}(0)$, and
$\nu_{\mathrm{ou}}(t) := \beta k_0 D_{\mathrm{ou}}(t)$ is the inverse of a
fluctuating relaxation time [For simplicity, however, we refer to
$\nu_{\mathrm{ou}}(t)$ also as the fluctuating diffusivity].
%% Due to the second term on the right hand side of
%% Eq.~(\ref{e.r(t)-R(t).clefd}), $\Delta\bm{r}(t)$ is a non-Gaussian process
%% \cite{uneyama19}; For a given path $D_{\mathrm{ou}}(u)$ [or, equivalently,
%% $\nu_{\mathrm{ou}}(u)$], however, $\Delta\bm{r}(t)$ is a Gaussian
%% process. Only if we take into account many realizations of the diffusivity
%% paths $[D_{\mathrm{ou}}(u)]$, $\Delta\bm{r}(t)$ becomes a non-Gaussian
%% process. It follows that $\bm{r}(t)$ is also a Gaussian process for each
%% realization $D_{\mathrm{ou}}(u)$; thanks to this property, the ISF can be
%% easily derived as shown below.

%subsection {GLE}
Substituting Eq.~(\ref{e.r(t)-R(t).clefd}) into Eq.~(\ref{e.dimer.dr(t)/dt}), we
have a Langevin equation without an integral term
\begin{equation}
  \label{e.gle}
  \frac {d \bm{r}(t)}{dt}
  =
  \sqrt{2D} \bm{\xi}_{\mathrm{all}}(t). 
\end{equation}
{\color{black}It is worth noting that the fluctuating diffusivity can be seen
  explicitly in this equation as explained shortly. Although such an equation
  without an integral term cannot be derived in general for the GLEFD
  [Eq.~(\ref{e.glefd})] due to the non-convolution form of the integral term, 
  Eq.~(\ref{e.gle}) demonstrates an essential role of the stochastic memory
  kernel $\phi(t,t')$ in the fluctuating diffusivity.}

In Eq.~(\ref{e.gle}), $\bm{\xi}_{\mathrm{all}}(t)$ is a correlated noise defined
by
\begin{align}
  \label{e.eta(t)}
  \bm{\xi}_{\mathrm{all}}(t)
  &=
  ~\bm{\xi}(t)
  -
  \sqrt{\frac {\nu\beta k_0}{2}}
  \Delta\bm{r}_0 e^{- \int_{0}^{t} \nu_{\mathrm{ou}}(u)du}
  \notag\\[0.1cm]
  &-
  \sqrt{\nu} \int_{0}^{t}dt_1
  \sqrt{\nu_{\mathrm{ou}}(t_1)}
  \bm{\xi}_{\mathrm{ou}}(t_1)  e^{-\int_{t_1}^{t} \nu_{\mathrm{ou}}(u)du},
\end{align}
where we define $\nu:= D\beta k_0$. From Eq.~(\ref{e.eta(t)}), the correlation
function of the noise $\bm{\xi}_{\mathrm{all}}(t)$ is given by
\begin{equation}
  \label{e.<eta(t)eta(t')>}
  \left\langle
  \bm{\xi}_{\mathrm{all}}(t) \bm{\xi}_{\mathrm{all}}(t')
  \right\rangle
  =
  \delta(t-t') \bm{I}
  -
  \frac {\nu}{2}
  e^{-\left|\int^{t}_{t'} \nu_{\mathrm{ou}}(u)du \right|} \bm{I}.
\end{equation}
Thus, this correlation function depends on the diffusivity path
$\nu_{\mathrm{ou}}(t)$.

If $D_{\mathrm{ou}}(t)$ and $\nu_{\mathrm{ou}}(t)$ are independent of time $t$,
the third term of the right hand side of Eq.~(\ref{e.eta(t)}) gives a correlated
Gaussian noise with an exponential correlation. However,
$\sqrt{\nu_{\mathrm{ou}}(t_1)}$ in the integrand of Eq.~(\ref{e.eta(t)})
indicates that the diffusivity is fluctuating, and the exponential correlation
is also influenced by the fluctuating diffusivity through
$\nu_{\mathrm{ou}}(u)$.  Thus, the fluctuating diffusivity
$\nu_{\mathrm{ou}}(t)$ and the correlated noise $\bm{\xi}_{\mathrm{all}}(t)$ are
coupled in a complicated way; it is contrasting to the models studied in
Refs.\cite{wang20, wang20b, sabri20}, in which the fluctuating diffusivity and
the correlated noise are assumed to be independent {\color{black}(Such a noise is
  briefly discussed in Sec.~\ref{s.discussion})}.

\subsection {Mean square displacement}\label{s.dimer.MSD}
%subsubsection {intro}

In this subsection, we derive formulas for the ensemble-averaged and
time-averaged MSDs. It shall be shown that the two MSDs coincide if the
fluctuating diffusivity $D_0(t)$ is a stationary process. A useful approximation
for the MSD is also derived under the assumption $D_0(t) \ll D$.

\subsubsection {Ensemble-averaged MSD}
Let us define a displacement vector $\delta \bm{r}(t)$ as
$\delta \bm{r}(t):= \bm{r}(t) - \bm{r}(0)$. From Eqs.~(\ref{e.gle}) and
(\ref{e.<eta(t)eta(t')>}), the ensemble-averaged MSD can be calculated as
\begin{equation}
  \label{e.emsd.1}
  \frac {\left\langle \delta\bm{r}^2(t) \right\rangle}{2n}
  =
  D
  \left[
  t
  - 
  \int_0^t dt_1 \int_0^{t_1} dt_2
  f[t_1,t_2 |\nu_{\mathrm{ou}}]
  %e^{-\kappa \int_{s'}^{s} \widetilde{\Lambda}(u)du}
  \right],
\end{equation}
where $f[t_1,t_2 | \nu_{\mathrm{ou}}]$, a functional of $\nu_{\mathrm{ou}}(t)$,
is defined by
\begin{align}
  f[t_1,t_2 | \nu_{\mathrm{ou}}]
  &=
  \nu\,
  e^{-\int_{t_2}^{t_1} \nu_{\mathrm{ou}}(u)du}
  \notag\\[0.1cm]
  \label{e.f[s,s'|D]}
  &=
  \nu\,
  e^{-\nu (t_1-t_2)}
  e^{-\int_{t_2}^{t_1} \nu_0(u)du},
\end{align}
for $t_2 < t_1$.  In the second equality,
$\nu_{\mathrm{ou}}(t) = \nu + \nu_0(t)$ is used.

Taking the ensemble average over the fluctuating diffusivity
$\nu_{\mathrm{ou}}(t)$ in Eq.~(\ref{e.emsd.1}), we have
\begin{align}
  \label{e.emsd.2}
  \frac {\dbra{\delta \bm{r}^2(t)}}{2n}
  &=
  D
  \left[
  t
  -
  \int_0^t dt_1 \int_0^{t_1} dt_2
  f(t_1,t_2 )
  \right],
\end{align}
where $\left\langle \dots \right\rangle_\nu$ denotes the ensemble average over
$\nu_0(t)$. The two-time function $f(t_1,t_2 )$ is defined as
\begin{equation}
  \label{e.f(s,s')}
  f(t_1,t_2 )
  =
  \nu\,
  e^{-\nu (t_1-t_2 )}
  \left\langle  e^{-\int_{t_2}^{t_1} \nu_{0}(u)du} \right\rangle_\nu.
\end{equation}
The ensemble average in the right-hand side of Eq.~(\ref{e.f(s,s')}) is the
normalized position correlation function for the Ornstein-Uhlenbeck process with
fluctuating diffusivity studied in Refs.~\cite{uneyama19, miyaguchi19}.
If $\nu_0(t)$ is a stationary process, Eq.~(\ref{e.emsd.2}) is further rewritten
as
\begin{equation}
  \label{e.emsd.stationary}
  \frac {\dbra{\delta \bm{r}^2(t)}}{2n}
  =
  D
  \left[
  t
  - 
  \int_{0}^{t} du
  (t-u) f(u)
  \right],
\end{equation}
where we define $f(t_1-t_2):= f(t_1,t_2 )$, and thus
\begin{equation}
  \label{e.f(s)}
  f(t)
  =
  \nu\,
  e^{-\nu t}
  \left\langle  e^{-\int_{0}^{t} \nu_{0}(u)du} \right\rangle_\nu.
\end{equation}

The equation (\ref{e.emsd.stationary}) is an exact formula for the case in which
$D_0(t)$ [or $\nu_0(t)$] is stationary. But, it is also possible to derive a
useful formula, which is valid even for non-stationary cases, if we assume
$D_0(t) \ll D$, or equivalently, $\nu_0(t) \ll \nu$ for any $t$. Under this
assumption, Eq.~(\ref{e.f[s,s'|D]}) can be approximated by using a functional
Taylor expansion \cite{binney92} as
\begin{equation}
  \label{e.f[s,s'|D].app}
  f[t_1,t_2 | \nu_{\mathrm{ou}}]
  \simeq
  \nu\,e^{- \nu  (t_1-t_2)}
  \left[1 - \int_{t_2}^{t_1}  \nu_0 (u) du\right].
\end{equation}
Substituting this equation in Eq.~(\ref{e.emsd.1}) and using integration by
parts, we obtain
\begin{align}
  \frac {\left\langle \delta \bm{r}^2(t)  \right\rangle}{2n}
  &\simeq
  \frac {1}{\beta k_0}
  \left[
  1
  +
  \int_0^tdu (1-e^{-\nu u}) \nu_0(u)
  \right]
  \notag\\[0.1cm]
  \label{e.msd.app}
  &-
  \frac {e^{- \nu t}}{\beta k_0} 
  \left[ 1 + \int_0^t du (e^{\nu u} - 1) \nu_0 (u) \right].
\end{align}

Neglecting terms of order $O(\nu_0(t)/\nu)$, we have a useful expression for the
MSD as
\begin{equation}
  \label{e.msd.app.overall}
  \frac {\left\langle \delta \bm{r}^2(t) \right\rangle}{2n}
  \simeq
  %\frac {1}{u} \left(1 - e^{-D_sut}\right) + \int_0^t ds D_l(s).
  D_s(t)\,t + \int_0^t du D_0(u),
\end{equation}
{\color{black}where $D_s(t)$ is a short-time diffusivity defined by
  \begin{equation}
    \label{e.D_s(t)}
    D_s(t) := (1 - e^{- \beta k_0 D t})/\beta k_0 t.
  \end{equation}
  Therefore, the short-time diffusivity is not fluctuating. Note also that
  $D_s(t)$ decays to zero as $t \to \infty$. Moreover, the second term in
  Eq.~(\ref{e.msd.app.overall}) contributes to the long-time diffusivity; thus,
  the long-time diffusivity is fluctuating. } As shown in figures in subsequent
sections, this formula for the MSD is remarkably consistent with numerical
simulations.

Taking the average of Eq.~(\ref{e.msd.app.overall}) over the fluctuating
diffusivity, we have
\begin{equation}
  \label{e.msd.app.overall.D}
  \frac {\dbra{\delta \bm{r}^2(t)}}{2n}
  \simeq
  %\frac {1}{u} \left(1 - e^{-D_sut}\right)
  D_s(t)\,t
  +
  \int_0^t du \left\langle D_0(u)  \right\rangle_\nu.
\end{equation}
It follows from Eq.~(\ref{e.msd.app.overall.D}) that the MSD shows a plateau
($\simeq 1/\beta k_0$) at an intermediate time scale
$1/\nu \ll t \ll 1/\nu_0(u)$.

If the fluctuating diffusivity $D_0(u)$ is a stationary process,
$\left\langle D_0(u) \right\rangle_\nu$ is independent of time $u$. Therefore,
denoting this as $D_0^{\mathrm{eq}} := \left\langle D_0(u) \right\rangle_\nu $,
we have from Eq.~(\ref{e.msd.app.overall.D})
\begin{equation}
  \label{e.msd.app.overall.D.steady}
  \frac {\dbra{\delta \bm{r}^2(t)}}{2n}
  \simeq
  \left[D_s(t)
  +
  D_0^{\mathrm{eq}}\right] t.
\end{equation}
Thus, $D_0^{\mathrm{eq}}$ is the long-time diffusivity, because $D_s(t) \to 0$
as $t \to \infty$. Moreover, Eq.~(\ref{e.msd.app.overall.D.steady}) shows that,
for any stationary diffusivity process $D_0(t)$, the long-time diffusion is
normal. In other words, statistical properties of the fluctuating diffusivity
$D_0(t)$ such as its correlation functions cannot be determined only by the MSD.

\subsubsection {Time-averaged MSD}

A time-averaged MSD is a method to obtain the MSD from time series, and it is
frequently used in experiments \cite{golding06, parry14}. The time-averaged MSD
is defined by
\begin{equation}
  \label{e.tamsd.def}
  \overline{\delta \bm{r}^2}(\Delta; t)
  :=
  \frac {1}{t} \int_{0}^{t} dt'
  \left[\bm{r}(t'+\Delta) - \bm{r}(t')\right]^2,
\end{equation}
where $\Delta$ is a lag time, and $t+\Delta$ is the length of the whole time
series.

Using Eq.~(\ref{e.gle}) in Eq.~(\ref{e.tamsd.def}), and taking an average over
the noise history $\bm{\xi}_{\mathrm{all}}(t)$ with
Eq.~(\ref{e.<eta(t)eta(t')>}), we obtain
\begin{equation}
  \label{e.<tmsd>}
  \frac {\left\langle
    \overline{\delta \bm{r}^2}(\Delta; t)
    \right\rangle}{2n}
  =
  D
  \left[
  \Delta -
  \int_{0}^{t} \frac {dt'}{t}
  \int_{t'}^{t'+\Delta}\!\!\!\!\!\!\!du
  \int_{t'}^{u}du' f[u, u'|\nu_{\mathrm{ou}}]
  \right]
\end{equation}
Moreover, taking an average over the fluctuating diffusivity
$\nu_{\mathrm{ou}}(t)$, we have
\begin{equation}
  \label{e.etmsd}
  \frac {\dbra{\overline{\delta \bm{r}^2}(\Delta; t)}}{2n}
  =
  D
  \left[
  \Delta -
  \int_{0}^{t} \frac {dt'}{t}
  \int_{t'}^{t'+\Delta}\!\!\!\!\!\!\!du
  \int_{t'}^{u}du' f(u, u')
  \right]
\end{equation}
Apparently, the time-averaged MSD [Eq.~(\ref{e.etmsd})] does not coincide with
the ensemble-averaged MSD [Eq.~(\ref{e.emsd.2})] in general.

However, if the fluctuating diffusivity $D_0(t)$ is stationary, we have
$f(u,u') = f(u-u')$. Consequently, Eq.~(\ref{e.etmsd}) is rewritten as
\begin{equation}
  \label{e.tmsd.stationary}
  \frac {\dbra{\overline{\delta \bm{r}^2}(\Delta; t)}}{2n}
  =
  D
  \left[
  \Delta -
  \int_{0}^{\Delta}du (\Delta - u) f(u)
  \right].
\end{equation}
Thus, for stationary $D_0(t)$, the two MSDs coincide
[Eqs.~(\ref{e.emsd.stationary}) and (\ref{e.tmsd.stationary})].

{\color{black}As in the ensemble-averaged MSD, a useful formula for the
  time-averaged MSD, which is valid even for non-stationary cases, can be
  derived under the assumption $D_0(t) \ll D$. Substituting
  Eq.~(\ref{e.f[s,s'|D].app}) in Eq.~(\ref{e.<tmsd>}) and using integration by
  parts, we obtain
  \begin{align}
    \frac {\left\langle \overline{\delta \bm{r}^2}(\Delta; t)  \right\rangle}{2n}
    &\simeq
    \frac {1}{\beta k_0}
    \left[
    1
    +
    \int_0^{\Delta}du (1-e^{-\nu u}) \overline{\nu}_0(u; t)
    \right]
    \notag\\[0.1cm]
    \label{e.tmsd.app}
    &-
    \frac {e^{- \nu \Delta}}{\beta k_0} 
    \left[ 1 + \int_0^{\Delta} du (e^{\nu u} - 1) \overline{\nu}_0 (u; t) \right],
  \end{align}
  where $\overline{\nu}_0(u; t)$ is a time average of $\nu_{0}(u)$ defined by
  $\overline{\nu}_0(u; t) := \int_0^t dt' \nu_0(u+t')/t$.
  
  Neglecting terms of order $O(\nu_0(t)/\nu)$, we have an approximated formula
  for the time-averaged MSD as
  \begin{equation}
    \label{e.tmsd.app.overall}
    \frac {\left\langle \overline{\delta \bm{r}}^2(\Delta;t) \right\rangle}{2n}
    \simeq
    D_s(\Delta)\,\Delta + \int_0^{\Delta} du \overline{D}_0(u; t),
  \end{equation}
  where $\overline{D}_0(u; t)$ is a time average defined in the same way as
  $\overline{\nu}_0(u; t)$.  Taking the average of
  Eq.~(\ref{e.tmsd.app.overall}) over the fluctuating diffusivity, we have
  \begin{equation}
    \label{e.tmsd.app.overall.D}
    \frac {\dbra{\overline{\delta \bm{r}}^2(\Delta; t)}}{2n}
    \simeq
    %\frac {1}{u} \left(1 - e^{-D_sut}\right)
    D_s(\Delta)\,\Delta
    +
    \int_0^{\Delta} du \left\langle \overline{D}_0(u; t)  \right\rangle_\nu,
  \end{equation}
  where $\left\langle \overline{D}_0(u; t) \right\rangle_{\nu}$ is given by
  \begin{equation}
    \label{e.tmsd.<D0(u;t)>_nu}
    \left\langle \overline{D}_0(u; t) \right\rangle_{\nu}
    =
    \int_0^t \frac {dt'}{t}
    \left\langle D_0(u+t')  \right\rangle_{\nu}.
  \end{equation}
  Apparently, Eq.~(\ref{e.tmsd.app.overall.D}) corresponds to the similar
  formula for the ensemble-averaged MSD in Eq.~(\ref{e.msd.app.overall.D}).  }

\subsection {Self-intermediate scattering function}
%subsubsection {intro}

It is possible to derive exact formulas for higher order moments, but resulting
equations become much more complicated than those for the MSD
[Eqs.~(\ref{e.emsd.2}) and (\ref{e.emsd.stationary})]. Instead of this approach,
we assume a separation of two time scales $1/\nu \ll 1/\nu_0(t)$ for any $t$
[i.e., $\nu_0 (t)\ll \nu$ or equivalently $D_0(t) \ll D$] and derive a formula
for the ISF.

Under the assumption $D_0(t) \ll D$, we derive a formula for the
ensemble-averaged MSD for each realization of the diffusivity $D_0(t)$
[Eq.~(\ref{e.msd.app.overall})]. As noted at the end of Sec.~\ref{s.gle}, the
GLEFD for a given path $D_0(t)$ is a Gaussian process. It follows that the ISF
$F_s[k, t|D_0]$ for the given path $D_0(t)$ is simply given by
\cite{miyaguchi19}
\begin{align}
  F_s[k, t | D_0]
  =&
  \exp\left[
  -k^2 \frac {\left\langle \delta \bm{r}^2(t) \right\rangle}{2n} 
  \right]
  \notag\\[0.1cm]
  \label{e.Fs[k,t|D]}
  \simeq&
  \exp\left[- 
  k^2D_s(t)t
  -
  k^2\int_{0}^{t} D_0(u)du
  \right].
\end{align}
Here, $F_s[k, t | D_0]$ is a Fourier transform of the displacement distribution
$G[\delta\bm{r}, t| D_0]$ for the given path $D_0(t)$, i.e.,
$F_s[k, t|D_0] := \int_{-\infty}^{\infty}d\delta\bm{r} e^{-i \bm{k}\cdot
  \delta\bm{r}} G[\delta\bm{r}, t|D_0]$ with $k := |
\bm{k}|$. {\color{black}Moreover, the time variable $t$ shall be Laplace
  transformed into a Laplace variable $s$ in the following section.}

Taking the average over the fluctuating diffusivity $D_0(t)$ [let us use the
same notation $\left\langle ... \right\rangle_\nu$ as before], we have the ISF as
\begin{align}
  F_s(k, t)
  \simeq&
  %\exp\left[- k^2 g(t) \right]
  e^{-k^2 D_s(t) t}
  \label{e.F(k,t)}
  \left\langle
  \exp\left[-k^2 \int_{0}^{t} D_0(u)du\right]
  \right\rangle_\nu.
\end{align}
The ensemble average in the right-hand side of Eq.~(\ref{e.F(k,t)}) is a
relaxation function thoroughly studied in Refs.~\cite{uneyama19} and
\cite{miyaguchi19}.
By using Eq.~(\ref{e.F(k,t)}) in
$\left\langle\!\left\langle \delta \bm{r}^2(t) \right\rangle\!\right\rangle_\nu
= -\nabla_k^2 F_s(k,t)|_{k=0}$, we can easily obtain
Eq.~(\ref{e.msd.app.overall.D}), where $\nabla_k$ is the gradient operator in
terms of $\bm{k}$. As in the case of the MSD, the ISF shows a plateau at an
intermediate timescale $1/\nu \ll t \ll 1/\nu_0(u)$, if the wave number $k$
satisfies $k^2/\beta k_0 \sim O(1)$.

\subsection {Non-Gaussian parameter}

To derive a formula for the NGP $A(t)$ [Eq.~(\ref{e.ngp-def})], the fourth order
moment $\dbra{\delta \bm{r}^4}$ should be obtained, but it can be readily
calculated with Eq.~(\ref{e.F(k,t)}) and
$\dbra{\delta \bm{r}^4} = \nabla_k^2\nabla_k^2 F_s(k,t)|_{k=0}$. The resulting
formula for $A(t)$ is
\begin{equation}
  \label{e.ngp}
  A(t)
  =
  \frac
  {\int_0^t du_1\int_0^t du_2 \left\langle \delta D_0(u_1)\delta D_0(u_2) \right\rangle_\nu}
  {
    \left[D_s(t)t +  \int_0^t \left\langle D_0(u)\right\rangle_\nu du \right]^2
  },
\end{equation}
%% \begin{widetext}
%%   \begin{equation}
%%     \label{e.ngp}
%%     A(t)
%%     =
%%     \frac
%%     {\int_0^t ds_1\int_0^t ds_2 \left\langle \delta D_0(s_1)\delta D_0(s_2) \right\rangle_\nu}
%%     {\int_0^t ds_1\int_0^t ds_2
%%     \left[\left\langle D_0(s_1)\right\rangle_\nu + D_s(s_1)\right]
%%     \left[\left\langle D_0(s_2)\right\rangle_\nu + D_s(s_2)\right]},
%%   \end{equation}
%% \end{widetext}
where $\delta D_0(t) := D_0(t) - \left\langle D_0(t) \right\rangle_\nu$.
Consequently, the NGP $A(t)$ can be utilized as a tool to elucidate
autocorrelation of the fluctuating diffusivity.  Equation (\ref{e.ngp}) is
almost the same as the quantity called the ergodicity breaking parameter studied
in Ref.~\cite{miyaguchi16}.

\section {Dimer model with two state diffusivity}\label{s.two-state-model}
%subsection {intro}

In this section, we study the dimer model under an assumption that the
diffusivity $D_0(t)$ of the auxiliary variable $\bm{r}_0(t)$ takes only two
values $D_+$ and $D_-$ as
\begin{align}
  \label{e.D0(t).two-state}
  D_0(t) :=
  \begin{cases}
    D_+ & (+ \text{state}; \text{fast state}),\\
    D_- & (- \text{state}; \text{slow state}).
  \end{cases}
\end{align}
{\color{black}Such a two-state model might well be plausible to describe
  tagged-particle motion in supercooled liquids and glasses, in which clusters
  of fast and slow particles are formed \cite{weeks00, miyaguchi16}.} We employ
the above notation instead of Eq.~(\ref{e.glefd.two-state}) just for consistency
with previous works \cite{miyaguchi16, miyaguchi19}. We assume $D_- < D_+$ and
refer to $D_+$ ($D_-$) as a fast (slow) state.  The diffusivity $D_0(t)$ is
assumed to switch between the fast and slow states at random times $t_1, t_2,
\dots$ in the same way as $\nu_i(t)$ in Eq.~(\ref{e.glefd.two-state}). The
sojourn-time distributions of the fast and slow states are again denoted as
$\rho^{\pm}(\tau)$.
%% at random times $t_1, t_2,\dots$. Let
%% $\tau_k := t_k - t_{k-1} \,(k=1,2,\dots)$ be sojourn times of the two states,
%% where we define $t_0 = 0$ for convenience.
%

In the following, $\rho^+(\tau)$ is assumed to follow the exponential
distribution with mean $\mu_+$ [Eq.~(\ref{e.exp-dist})], and $\rho^-(\tau)$ to
follow either the exponential distribution with mean $\mu_-$ or a power-law
distribution with index $\alpha$, i.e., $\rho^-(\tau) \sim \tau^{-1-\alpha}$
[Eq.~(\ref{e.rho(t)})]. For the power law with $1 < \alpha < 2$, we focus on an
equilibrium ensemble; for $0 < \alpha < 1$, however, the system does not reach
an equilibrium state, and thus we employ a typical non-equilibrium initial
ensemble \cite{miyaguchi16, miyaguchi19}.

The power-law sojourn times should be important for a heterogeneous diffusion
process. For example, suppose that a medium inside which the particle diffuses
is composed of fast and slow regions; in the fast (slow) region, the particle
diffuses with $D_+$ ($D_-$). In such a case, the particle visits one region
after another, and the sojourn time in a region for each visit might follow a
power law with an exponential cutoff. 

{\color{black}The two-state dimer model is related to the continuous-time random
  walk \cite{metzler00}. In fact, under the conditions that $D_-=0$ as well as
  $D_+\to \infty$ with $D_+ \mu_+$ being fixed, behavior of the two-state dimer
  model is similar to that of the continuous-time random walk (See
  Ref.~\cite{uneyama15} for detail). More precisely, with the above conditions,
  the two-state dimer model can be considered as a model of tagged-particle
  motion harmonically coupled with a heavy CTRW particle. Even without the above
  conditions, however, the two-state dimer model still shares many properties
  such as the non-Gaussianity with the continuous-time random walk
  \cite{miyaguchi16}.}
  
\subsection {Exponential distribution}\label{s:dimer-exp}
%subsubsection {fig3}
\begin{figure}[t!]
  \centerline{\includegraphics[width=\fsize]{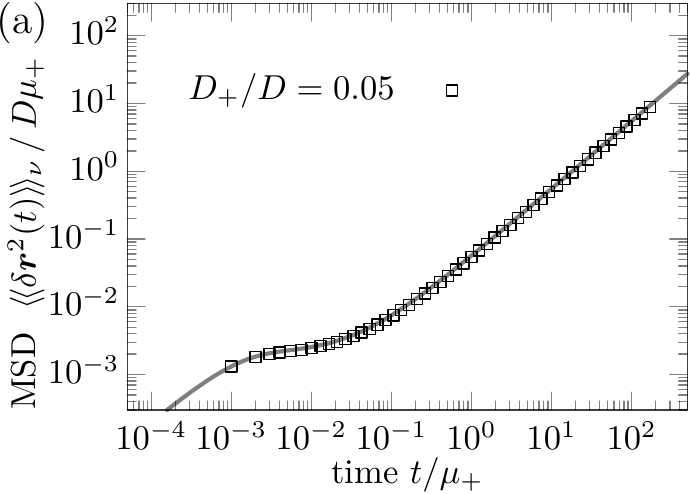}}
  \centerline{\includegraphics[width=\fsize]{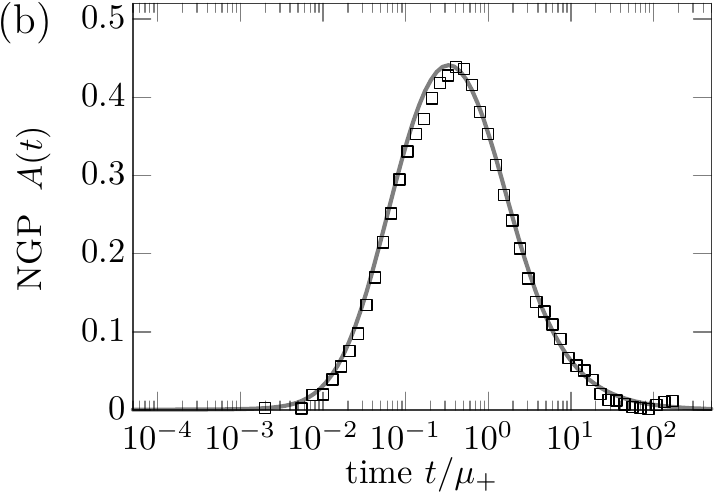}}
  \centerline{\includegraphics[width=\fsize]{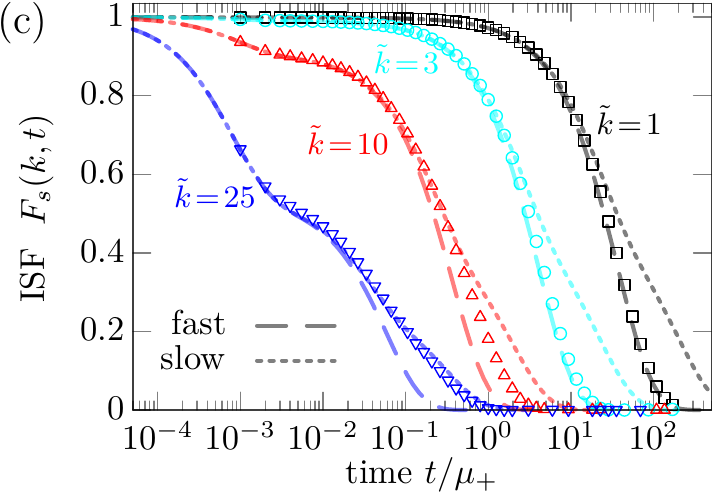}}
  \caption{\label{f.exp-dist} Equilibrium dimer model [$\rho^{\pm}(\tau)$ are
    both exponential distributions]. The symbols are data obtained by numerical
    integration of Eqs.~(\ref{e.dimer.dr(t)/dt}) and
    (\ref{e.dimer.dr1(t)/dt}). The fast and slow state diffusivities $D_\pm$ are
    set as $D_+/D = 0.05$ and $D_- = 0.1D_+$, the potential parameter $k_0$ as
    $\beta k_0 D \mu_+= 10^3$, the mean sojourn time $\mu_-$ as $\mu_- =
    \mu_+$. (a) MSD vs time.  The line is the theoretical prediction in
    Eq.~(\ref{e.msd.app.overall.D.steady}). (b) NGP vs time. The line is the
    prediction in Eq.~(\ref{e.ngp.exp}). (c) ISF vs time.  The lines are the
    predictions in Eqs.~(\ref{e.Fs(k,t).fast-switch.2}) (the dashed lines) and
    (\ref{e.Fs(k,t).slow-switch}) (the dotted lines). From right to left, the
    wave number $k$ is set as $k\sqrt{D\mu_+} = 1, 3, 10, 25\, (=: \tilde{k})$.  }
\end{figure}

%subsubsection {theory 1}
First, let us investigate the case in which the sojourn-time distributions
$\rho^{\pm}(\tau)$ are given by the exponential distributions with mean sojourn
times $\mu_{\pm}$. Then, equilibrium fractions of the two states are given by
\begin{equation}
  \label{e.eq-ensemble-1}
  p_{\pm}^{\mathrm{eq}}
  =
  \frac {\mu_{\pm}}{\mu_+ + \mu_-}.
  %% = \frac {k_{\mp}}{k_+ + k_-}
  %% \frac {\mu_{\pm}}\mu
\end{equation}
Initially, the particle is in the fast or the slow state according to the
fractions $p_{\pm}^{\mathrm{eq}}$. Then, the Laplace transform of the relaxation
function [Eq.~(\ref{e.F(k,t)})] is given by [Eq.~(46) in
Ref.~\cite{miyaguchi19}]
\begin{equation}
  \mathcal{L}\left[
  \left\langle  e^{-k^2 \int^t_0 D_0(u)du} \right\rangle_\nu
  \right](s)
  \label{e.relaxation_func.exp-dist.laplace}
  =
  \frac {z_+/\mu_+ + z_-/\mu_- + \mathcal{K}^2}
  {\mathcal{K}^2[s + D_0^{\mathrm{eq}}k^2 +z_+z_-/\mathcal{K}]}.
\end{equation}
where $\mathcal{K} := \mu_+^{-1}+\mu_-^{-1}$, and $z_{\pm}:= s +
D_{\pm}k^2$. $\mathcal{L}[\dots]$ represents the Laplace transform defined by
$\mathcal{L}[g(t)](s) :=\int_0^{\infty}g(t)e^{-st}dt$.  The equilibrium
diffusivity $D_0^{\mathrm{eq}}$ is here calculated as
$D_0^{\mathrm{eq}} = p_+^{\mathrm{eq}} D_+ + p_-^{\mathrm{eq}} D_-$.

%subsubsection {Fs as k to 0}

For a derivation of the ensemble-averaged MSD and the NGP, a small $k$ limit of
the ISF $F_s(k, t)$ can be utilized. After a straightforward but slightly
lengthy calculation, we obtain
\begin{align}
  \mathcal{L}&\left[
  \left\langle  e^{-k^2 \int^t_0 D_0(u)du} \right\rangle_\nu
  \right](s)
  \notag\\[0.1cm]
  \label{e.relaxation_func.exp-dist.laplace.approx.at.k=0}
  &\underset{k\to0}{\simeq}
  \frac {1}{s}
  - \frac {D_0^{\mathrm{eq}}k^2}{s^2} 
  + \frac {(D_0^{\mathrm{eq}}k^2)^2}{s^3}
  + \frac {p_+^{\mathrm{eq}} p_-^{\mathrm{eq}}(\Delta D_0)^2}
  {s^2(s+ \mathcal{K})} k^4,
\end{align}
where $\Delta D_0 := D_+- D_-$. Plugging the Laplace inversion of
Eq.~(\ref{e.relaxation_func.exp-dist.laplace.approx.at.k=0}) into
Eq.~(\ref{e.F(k,t)}), we obtain the ISF
\begin{align}
  F_s&(k,t)
  \underset{k \to 0}{\simeq}
  \notag\\[0.1cm]
  \label{e.Fs(k,t).fast-switch}
  &\exp\left\{
  -k^2[D_s(t) + D_0^{\mathrm{eq}}]t
  + \frac {k^4p_+^{\mathrm{eq}} p_-^{\mathrm{eq}}}{2} 
  (\Delta D_0t)^2g_{d}(\mathcal{K} t)
  \right\},
\end{align}
where $g_d(t)$ is the Debye function defined by \cite{doi86}
\begin{equation}
  \label{e.debye-func}
  g_d(t) :=
  \frac {2}{t^2}
  \left(e^{-t} - 1 + t\right).
\end{equation}
Note that Eq.~(\ref{e.relaxation_func.exp-dist.laplace}) is an exact formula,
whereas Eq.~(\ref{e.Fs(k,t).fast-switch}) and Eq.~(\ref{e.F(k,t).exp-dist})
below are valid only when $D_{\pm} \ll D$, because Eq.~(\ref{e.F(k,t)}) was
used.

%subsubsection {msd}

Then, with the relation
$\left\langle\!\left\langle \delta \bm{r}^2(t) \right\rangle\!\right\rangle_\nu
= -\nabla_k^2 F_s(k,t)|_{k=0}$, it is easy to see that the ensemble-averaged MSD
is given by Eq.~(\ref{e.msd.app.overall.D.steady}) [or, it is evident because
Eq.~(\ref{e.Fs(k,t).fast-switch}) is a cumulant expansion]. As shown in
Fig.~\ref{f.exp-dist}(a), the prediction is consistent with the simulation. In
particular, the MSD exhibits a plateau at the intermediate timescale
$1/\nu \ll t \ll 1/\nu_0(u)$, which is
$10^{-3} \ll t/\mu_+ \ll 2 \times 10^{-2}$ for the parameter values employed in
Fig.~\ref{f.exp-dist}.

%subsubsection {ngp}
Similarly, the NGP $A(t)$ [Eq.~(\ref{e.ngp-def})] can be derived with
$\dbra{\delta \bm{r}^4} = \nabla_k^2\nabla_k^2 F_s(k,t)|_{k=0}$, and we obtain
\begin{equation}
  \label{e.ngp.exp}
  A(t) =
  \frac
  {
    p_+^{\mathrm{eq}}p_-^{\mathrm{eq}}(\Delta D_0)^2
    g_d(\mathcal{K}t)}
  {[D_s(t) + D_0^{\mathrm{eq}}]^2}.
\end{equation}
As shown in Fig.~\ref{f.exp-dist}(b), this prediction is also remarkably
consistent with numerical simulation.
Alternatively, it is also possible to derive Eq.~(\ref{e.ngp.exp}) with
Eq.~(\ref{e.ngp}). In fact, by using Eqs.~(17) and (66) in
Ref.~\cite{miyaguchi16}, we have the diffusivity correlation as
\begin{equation}
  \label{e.<deltaD(t)deltaD(0)>.exp-dist}
  \left\langle \delta D_0(t)\delta D_0(0) \right\rangle_\nu
  =
  (\Delta D_0)^2 p_+^{\mathrm{eq}} p_-^{\mathrm{eq}} e^{- \mathcal{K}t}.
\end{equation}
Then, substituting Eq.~(\ref{e.<deltaD(t)deltaD(0)>.exp-dist}) in
Eq.~(\ref{e.ngp}), we obtain Eq.~(\ref{e.ngp.exp}) again.

%subsubsection {isf}

Next, let us derive the ISF $F_s(k, t)$ without the assumption of $k$ being
small. To do this, the Laplace inversion of
Eq.~(\ref{e.relaxation_func.exp-dist.laplace}) is carried out as
\cite{uneyama19}
\begin{equation}
  \label{e.relaxation_func.exp-dist}
  \left\langle  e^{-k^2 \int^t_0 D_0(u)du} \right\rangle_\nu
  =
  A_+ e^{-\lambda_+ t} + A_- e^{-\lambda_- t},
\end{equation}
where the relaxation constants $\lambda_\pm$ are the simple poles of
Eq.~(\ref{e.relaxation_func.exp-dist.laplace}). $\lambda_\pm$ are given by
\begin{equation}
  \label{e.dimer.exp-dist.lambda}
  \lambda_{\pm} =  \mathcal{K}
  \frac {
    1 + \frac {D_+ + D_-}{\mathcal{K}}k^2
    \pm
    \sqrt{
      1
      - \frac {2k^2}{\mathcal{K}} \Delta p \Delta D_0
      + \left(\frac {k^2\Delta D_0}{\mathcal{K}}\right)^2
    }
  }{2},
\end{equation}
where $\Delta p := p_+^{\mathrm{eq}}-p_-^{\mathrm{eq}}$. Moreover, $A_{\pm}$,
the fractions of the two modes, are given by
\begin{equation}
  \label{e.dimer.exp-dist.A}
  A_{\pm} = \pm \frac {D^{\mathrm{eq}}_0k^2 - \lambda_{\mp}}{\lambda_+- \lambda_-}.
\end{equation}
Note that $A_+ + A_-=1$.  Then, by using Eq.~(\ref{e.F(k,t)}) and
(\ref{e.relaxation_func.exp-dist}), we have the ISF as
\begin{equation}
  \label{e.F(k,t).exp-dist}
  F_s(k, t)
  =
  A_+e^{-[k^2 D_s(t) + \lambda_+]t} +  
  A_-e^{-[k^2 D_s(t) + \lambda_-]t}. 
\end{equation}
{\color{black}It is possible to derive Eq.~(\ref{e.Fs(k,t).fast-switch}) by
  expanding Eq.~(\ref{e.F(k,t).exp-dist}) with small $k$ up to the fourth order
  $k^4$, thereby obtaining the MSD and NGP again.}
%subsubsection {isf: extreme}

There are two extreme cases which are interesting to study: a fast switching
limit and a slow switching limit.  In the fast switching limit, switching
between the two states is faster than the diffusion timescales $1/k^2D_{\pm}$,
i.e., $\mu_{\pm} \ll 1/k^2D_{\pm}$. This is equivalent to take $k \to 0$, and
thus from Eq.~(\ref{e.Fs(k,t).fast-switch}), we have
\begin{equation}
  \label{e.Fs(k,t).fast-switch.2}
  F_s(k,t)
  \underset{k\to0}{\simeq} \exp\left\{ -k^2[D_s(t) + D_0^{\mathrm{eq}}]t \right\}.
\end{equation}
Consequently, a single mode relaxation is observed for the fast switching limit
$k \to 0$.

The condition for the other extreme case, i.e., the slow switching limit, is
given by $\mu_{\pm} \gg 1/k^2D_{\pm}$. In this limit, from
Eqs.~(\ref{e.dimer.exp-dist.lambda}) and (\ref{e.dimer.exp-dist.A}), we have
$\lambda_{\pm} \to k^2D_{\pm}$ and $A_{\pm} \to p_{\pm}^{\mathrm{eq}}$. Thus, we
have from Eq.~(\ref{e.relaxation_func.exp-dist})
\begin{equation}
  \label{e.Fs(k,t).slow-switch}
  F_s(k,t)
  \underset{k\to\infty}{\simeq}
  p_+^{\mathrm{eq}}e^{-k^2[D_s(t) + D_+]t} +  
  p_-^{\mathrm{eq}}e^{-k^2[D_s(t) + D_-]t}. 
\end{equation}
This is a simple superposition of the two diffusion processes with diffusivities
$D_{\pm}$.

The important point is that, by changing the wave number $k$, the ISF varies
between these two extremes [Eqs.~(\ref{e.Fs(k,t).fast-switch.2}) and
(\ref{e.Fs(k,t).slow-switch})]. Therefore, the ISF might be a very useful tool
to elucidate the fluctuating diffusivity \cite{miyaguchi19, dieball22}. In
Fig.~\ref{f.exp-dist}(c), these two predictions
[Eq.~(\ref{e.Fs(k,t).fast-switch.2}) and (\ref{e.Fs(k,t).slow-switch})] are
compared with simulation. For small $k$, the fast switching limit (the dashed
lines) is consistent with the simulation, whereas, for large $k$, the slow
switching limit (the dotted lines) is consistent. Moreover, at intermediate
values of $k$ [for example, $\tilde{k}=10$ and $25$ in
Fig.~\ref{f.exp-dist}(c)], a two-step relaxation can be observed.

\subsection {Power law: Equilibrium ensemble ($1 < \alpha < 2$)}
%subsubsection {intro}

In this subsection, we assume that the the sojourn-time distribution of the fast state
$\rho^+(\tau)$ follows the exponential distribution with mean $\mu_+$, whereas
that of the slow state $\rho^-(\tau)$ follows a power-law with index
$1 < \alpha < 2$ and mean $\mu_-$ [Eq.~(\ref{e.rho(t)})]. The equilibrium
fractions of the two states are again given by
Eq.~(\ref{e.eq-ensemble-1}). Moreover, for the equilibrium ensemble, the first
sojourn time is chosen from the sojourn-time distribution $\rho^{\pm, \mathrm{eq}}(\tau)$
defined in Eq.~(\ref{e.app.rho^eq}) instead of $\rho^{\pm}(\tau)$ (See
Ref.~\cite{miyaguchi19} for detail).

Under these conditions, the Laplace transform of the relaxation function in
Eq.~(\ref{e.F(k,t)}) is given by [Eq.~(51) in Ref.~\cite{miyaguchi19}]
\begin{align}
  \mathcal{L}&\left[
  \left\langle  e^{-k^2 \int^t_0 D_0(u)du} \right\rangle_\nu
  \right](s)
  \notag\\[0.1cm]
  \label{e.relaxation_func.1<alpha<2.laplace}
  &\simeq
  \frac {1}{s + k^2D_0^{\mathrm{eq}}}
  +
  \frac {a(p_+^{\mathrm{eq}}\Delta D_0)^2 (s+k^2 D_-)^{\alpha-2} k^4}
  {\mu \left(s + k^2D_0^{\mathrm{eq}}\right)^2},
\end{align}
where $\mu := \mu_+ + \mu_-$, and $a$ is a constant characterizing the power law
[Eq.~(\ref{e.rho(t)})].  This asymptotic relation
[Eq.~(\ref{e.relaxation_func.1<alpha<2.laplace})] is valid if the conditions
$s \ll 1/\mu_\pm$ and $k^2 D_\pm \ll 1/\mu_\pm$ are both satisfied
\cite{miyaguchi19}; the latter condition is the fast switching assumption used
in the previous subsection.

%subsubsection {fig4}
\begin{figure}[t!]
  \centerline{\includegraphics[width=\fsize]{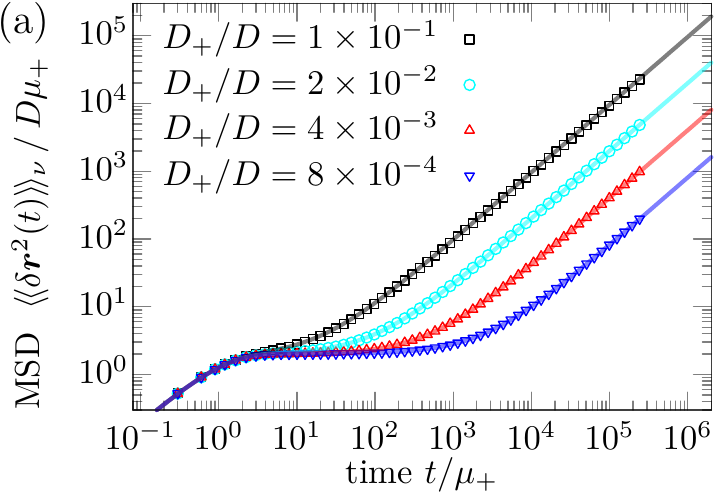}}
  \centerline{\includegraphics[width=\fsize]{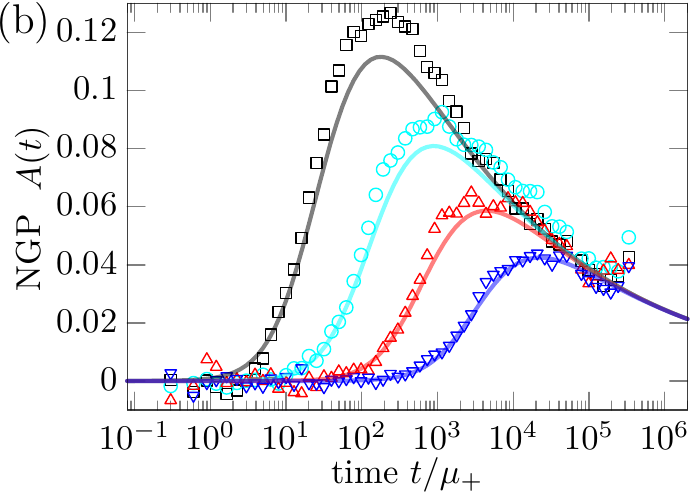}}
  \centerline{\includegraphics[width=\fsize]{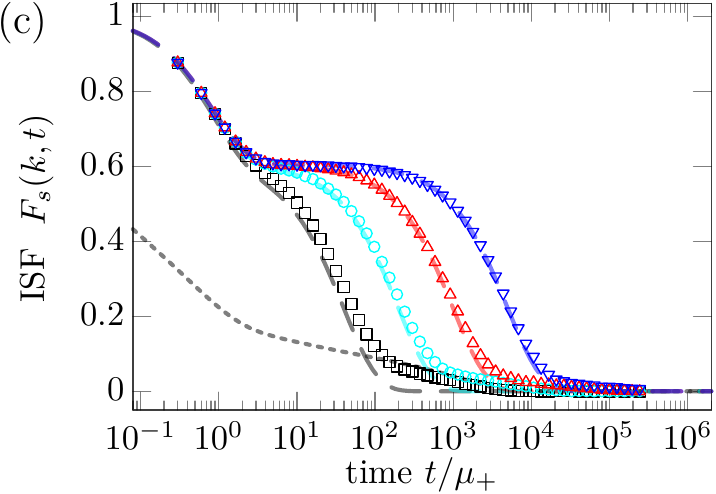}}
  \caption{\label{f.eq.alpha=115} Equilibrium dimer model [$\rho^{+}(\tau)$ and
    $\rho^{-}(\tau)$ are an exponential distribution and a power law,
    respectively]. The symbols are data obtained by numerical integration of
    Eqs.~(\ref{e.dimer.dr(t)/dt}) and (\ref{e.dimer.dr1(t)/dt}).  The fast and
    slow state diffusivities $D_\pm$ are set as
    $D_+/D = 10^{-1}$, $2\times 10^{-2}$, $4\times 10^{-3}$, $8\times 10^{-4}$, and
    $D_- = 0.01D_+$. The power-law index $\alpha$ is fixed as $\alpha = 1.2$,
    the potential parameter $k_0$ as $\beta k_0 D \mu_+= 1$, the mean sojourn
    time $\mu_-$ as $\mu_- = \mu_+$, and the cutoff parameter $\tau_0$ as
    $\tau_0 = \mu_-(\alpha-1)/\alpha$ [Eq.~(\ref{e.simulation.rho(tau)})]. (a)
    MSD vs time. The lines are the theoretical prediction in
    Eq.~(\ref{e.msd.app.overall.D.steady}). (b) NGP vs time. The lines are the
    prediction in Eq.~(\ref{e.ngp.eq}). (c) ISF vs time. The lines are the
    prediction in Eq.~(\ref{e.F(k,t).1<alpha<2}). The wave number $k$ is set as
    $k = \sqrt{\beta k_0/2}$.  [In (b-c), results for the four different values
    of $D_+$ are displayed with the same color code as in (a).].  }
\end{figure}

%subsubsection {fig5}
\begin{figure}[t!]
  \centerline{\includegraphics[width=\fsize]{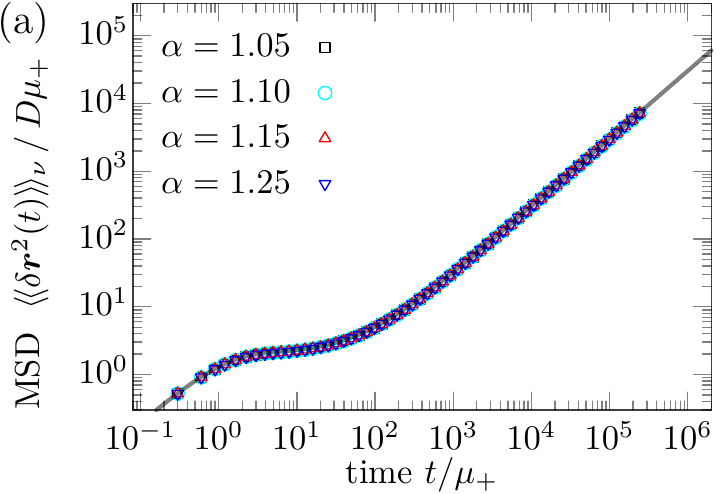}}
  \centerline{\includegraphics[width=\fsize]{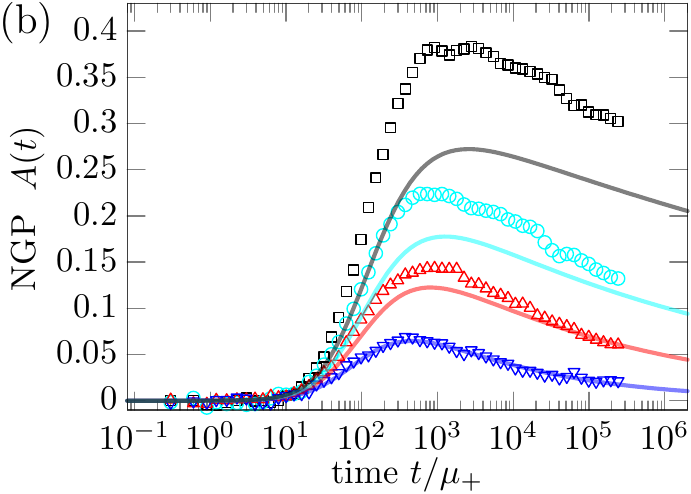}}
  \centerline{\includegraphics[width=\fsize]{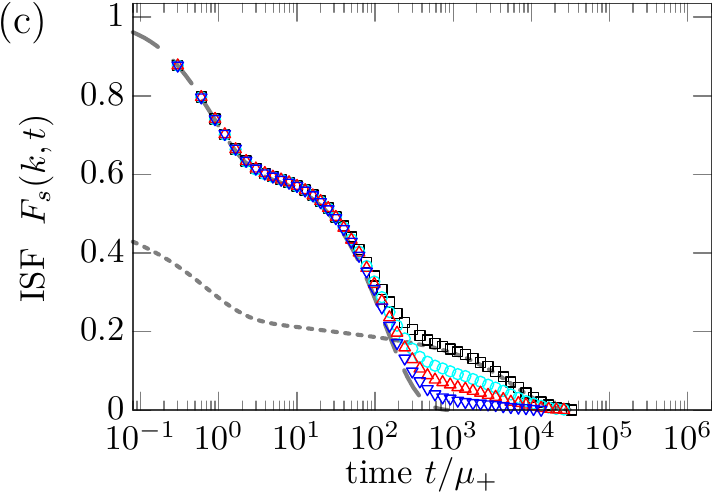}}
  \caption{\label{f.eq.df=01} Equilibrium dimer model (the same model as in
    Fig.~\ref{f.eq.alpha=115}). The power-law index is set as $\alpha = 1.05$,
    $1.1$, $1.15$ and $1.25$. The diffusivity $D_{\pm}$ of the two states are
    fixed as $D_+/D = 3.0 \times 10^{-2}$ and $D_- = 0.01 D_+$. Parameters
    $k_0$, $\mu_-$, $\tau_0$, and $k$ are the same as those in
    Fig.~\ref{f.eq.alpha=115}.  (a) MSD vs time. The line is the theoretical
    prediction in Eq.~(\ref{e.msd.app.overall.D.steady}).  (b) NGP vs time. The
    lines are the prediction in Eq.~(\ref{e.ngp.eq}). (c) ISF vs time. The lines
    are the prediction in Eq.~(\ref{e.F(k,t).1<alpha<2}). [In (b-c), results for
    the four different values of $\alpha$ are displayed with the same color code
    as in (a).}
\end{figure}

%subsubsection {msd, ngp, and isf}

As in the previous subsection, a small $k$ limit of the ISF $F_s(k, t)$ is
utilized to derive the ensemble-averaged MSD and the NGP.  Expanding
Eq.~(\ref{e.relaxation_func.1<alpha<2.laplace}) around $k = 0$, carrying out the
Laplace inversion, and inserting the resulting equation into
Eq.~(\ref{e.F(k,t)}), we obtain
\begin{align}
  \label{e.Fs(k,t)_small_k_eq}
  F_s&(k, t) \simeq e^{-k^2 D_s(t)t}
  \notag\\[0.1cm]
  \times&
  \left[
  1 - k^2{D_0^{\mathrm{eq}}}t + k^4 \frac {(D_0^{\mathrm{eq}})^2t^2}{2} 
  +
  k^4 \frac {(\Delta D_0p_+^{\mathrm{eq}})^2}{\Gamma(4-\alpha)}
  \frac {a}{\mu}t^{3-\alpha}  
  \right]
  \notag\\[0.1cm]
  \simeq&
  \exp\left\{
  -k^2 [D_s(t) + D_0^{\mathrm{eq}}]t
  + k^4 \frac {(\Delta D_0 p_+^{\mathrm{eq}})^2}{\Gamma(4-\alpha)}
  \frac {a}{\mu}t^{3-\alpha}
  \right\}.
\end{align}
Then, with the relation
$\left\langle\!\left\langle \delta \bm{r}^2(t) \right\rangle\!\right\rangle_\nu
= -\nabla_k^2 F_s(k,t)|_{k=0}$, it is easy to see that the ensemble-averaged MSD
is given by Eq.~(\ref{e.msd.app.overall.D.steady}). As shown in
Figs.~\ref{f.eq.alpha=115}(a) and \ref{f.eq.df=01}(a), the theoretical
prediction is consistent with the numerical simulations. In particular, the
ensemble-averaged MSD shows only normal diffusion and does not depend on the
power-law index $\alpha$ as exemplified in Fig.~\ref{f.eq.df=01}(a).

In contrast, non-Gaussianity depends on $\alpha$; in fact, the NGP $A(t)$
[Eq.~(\ref{e.ngp-def})] can be derived with
$\dbra{\delta \bm{r}^4} = \nabla_k^2\nabla_k^2 F_s(k,t)|_{k=0}$ as
\begin{equation}
  \label{e.ngp.eq}
  A(t) \simeq 
  \frac {(\Delta D_0\, p_+^{\mathrm{eq}})^2}{\Gamma(4-\alpha)}
  \frac {2a}{\mu} \frac {t^{1-\alpha}}{\left(1/\beta k_0t + D_0^{\mathrm{eq}}\right)^2},
\end{equation}
where the short-time diffusivity $D_s(t)$ is replaced with its asymptotic form
at long times $D_s(t) \simeq 1/\beta k_0t$, because
Eq.~(\ref{e.relaxation_func.1<alpha<2.laplace}) is valid only for large $t$
(i.e., $t \gg \mu_\pm$). The above formula [Eq.~(\ref{e.ngp.eq})] can be also
derived through Eq.~(\ref{e.ngp}) with the correlation function of diffusivity
[Eq.(111) in Ref.~\cite{miyaguchi16}]
\begin{equation}
  \label{e.<deltaD(t)deltaD(0)>}
  \left\langle \delta D_0(t)\delta D_0(0) \right\rangle_\nu
  \simeq
  \frac {(\Delta D_0\, p_+^{\mathrm{eq}})^2}{\Gamma(2-\alpha)}
  \frac {a}{\mu} t^{1-\alpha}.
\end{equation}

As shown in Fig.~\ref{f.eq.alpha=115}(b) and \ref{f.eq.df=01}(b), both of the
theory and simulations of the NGP $A(t)$ show unimodal structures, but around
the peaks, there are systematic {\color{black}deviations}; for $\alpha$ closer to
unity, the {\color{black}deviation} is more serious. Therefore, these
{\color{black}deviations} might be attributed to the higher order corrections
neglected in Eq.~(\ref{e.relaxation_func.1<alpha<2.laplace}).  At long times,
the theoretical prediction is consistent with the numerical simulations (as it
should be) except for the case $\alpha = 1.05$; for $\alpha = 1.05$, the
simulation time might well be shorter than the time scale at which the
asymptotic relation [Eq.~(\ref{e.ngp.eq})] becomes precise.

Finally, a formula for the ISF is derived without the assumption of $k$ being
small. The Laplace inversion of Eq.~(\ref{e.relaxation_func.1<alpha<2.laplace})
gives [Eqs.~(53) and (55) in Ref.~\cite{miyaguchi19}]
\begin{align}
  &\left\langle  e^{-k^2 \int^t_0 D_0(u)du} \right\rangle_\nu
  \notag\\[0.1cm]
  \label{e.relaxation_func.1<alpha<2}
  &\simeq
  \begin{cases}
    e^{-k^2 D_0^{\mathrm{eq}} t} & (t \sim \frac {1}{D_+k^2}),
    \\[0.1cm]
    \frac {a}{\mu}
    \left(\frac {p_+^{\mathrm{eq}} \Delta D_0}{D_0^{\mathrm{eq}}}\right)^2
    \frac {e^{- k^2 D_- t} t^{1-\alpha}}{{\color{black}|\Gamma(2-\alpha)|}}
    & (t \gg \frac {1}{D_+ k^2}).
  \end{cases}
\end{align}
By using Eqs.~(\ref{e.F(k,t)}) and (\ref{e.relaxation_func.1<alpha<2}),
the ISF is given by
\begin{equation}
  \label{e.F(k,t).1<alpha<2}
  F_s(k, t)
  \simeq
  \begin{cases}
    e^{-k^2 [D_0^{\mathrm{eq}} + D_s(t)]t} &
    (t \sim \frac {1}{D_+k^2}),
    \\[0.1cm]
    \frac {a}{\mu}
    \left(\frac {p_+^{\mathrm{eq}} \Delta D_0}{D_0^{\mathrm{eq}}}\right)^2
    \frac {e^{-k^2[D_s(t)+D_-]t} t^{1-\alpha}}{{\color{black}|\Gamma(2-\alpha)|}}
    & (t \gg \frac {1}{D_+k^2}).
  \end{cases}
\end{equation}
As show in Fig.~\ref{f.eq.alpha=115}(c) with dashed lines, the first of these
functions shows two-step relaxation. There is a plateau between the two
relaxation steps, and the duration of the plateau becomes longer for smaller
diffusivity $D_+$.

In Figs.~\ref{f.eq.alpha=115}(c) and \ref{f.eq.df=01}(c), the theoretical
prediction [the first equation in Eq.~(\ref{e.F(k,t).1<alpha<2})] is compared
with the numerical simulation. The theoretical prediction (the dashed lines) is
consistent with the numerical simulation (symbols) except at long times, at
which the second equation in Eq~.(\ref{e.F(k,t).1<alpha<2}) shows good agreement
[to avoid complication, the second equation is shown by dotted lines only for
the cases $D_+/D =0.1$ in Fig.~\ref{f.eq.alpha=115}(c) and $\alpha=1.05$ in
Fig.~\ref{f.eq.df=01}(c)]. In simulations, the wave number $k$ is set as
$k=\sqrt{\beta k_0/2}$, the inverse of which, $\sqrt{2/\beta k_0}$, is a
characteristic size of the harmonic potential since $2D_s(t)t \to 2/\beta k_0$
as $t \to \infty$ \cite{Fodor16}.

{\color{black} Finally, it is worth commenting that the formulas for the MSD
  [Eq.(\ref{e.msd.app.overall.D.steady})] and the ISF
  [Eq.~(\ref{e.F(k,t).1<alpha<2})] are valid even for the case $\alpha > 2$
  \cite{miyaguchi19}. In contrast, the NGP formula in Eq.~(\ref{e.ngp.eq}) is no
  more valid for $\alpha > 2$, and it should be necessary to include a leading
  term correctly.}

\subsection {Power law: Non-equilibrium ensemble ($0 < \alpha < 1$)} \label{s.dimer.noneq}
%subsubsection {intro}
Next, we investigate the case in which the sojourn time distribution
$\rho^-(\tau)$ is a power-law with $0 < \alpha < 1$. In this case, there is no
equilibrium state, because the mean sojourn time for this power-law distribution
does not exist. Therefore, we employ a typical non-equilibrium ensemble as an
initial ensemble, for which the first sojourn time simply follows $\rho^-(\tau)$
\cite{miyaguchi16, miyaguchi19}. The sojourn-time distribution for the fast
state $\rho^+(\tau)$ is again assumed to be the exponential distribution with
mean $\mu_+$. {\color{black}Furthermore, the initial fractions $p_{\pm}^0$
  ($p_+^0 + p_-^0=1$) for the two states are necessary for specifying the
  initial ensemble. However, the results obtained below for large $t$ are
  independent of $p_{\pm}^0$. }
 
Under these conditions, the Laplace transform of the relaxation function in
Eq.~(\ref{e.F(k,t)}) is given by [Eq.~(62) in Ref.~\cite{miyaguchi19}]
\begin{align}
  &\mathcal{L}\left[
  \left\langle  e^{-k^2 \int^t_0 D_0(u)du} \right\rangle_\nu
  \right](s)
  \simeq
  \notag\\[0.1cm]
  \label{e.relaxation_func.0<alpha<1.laplace}
  &
  \frac {1}{z_-}
  -
  \frac {\mu_+ k^2\Delta D_0}
  {az_-^{1+\alpha}}
  +
  \frac {\mu_+^2 k^2\Delta D_0z_+}
  {a^2z_-^{1+2\alpha}}
  +
  \frac {p_-^0 \mu_+ k^2\Delta D_0}{az_-},
\end{align}
where $z_{\pm} := s + k^2 D_{\pm}$. In Ref.~\cite{miyaguchi19}, the last two
terms on the right-hand side of Eq.~(\ref{e.relaxation_func.0<alpha<1.laplace})
are not taken into account, but terms up to $k^4$ are necessary to obtain the
NGP [The derivation procedure of Eq.~(\ref{e.relaxation_func.0<alpha<1.laplace})
is the same as that of Eq.(62) in Ref.~\cite{miyaguchi19}]. In a derivation of
Eq.~(\ref{e.relaxation_func.0<alpha<1.laplace}), the fast switching assumption
for the fast state $ k^2 D_+\ll 1/\mu_+$ as well as a similar condition for the
slow state $k^2D_- \ll 1/a^{1/\alpha}$ are assumed. Similarly, it is also
supposed that $s$ satisfies $s \ll 1/\mu_+$ and $s \ll 1/a^{1/\alpha}$
\cite{miyaguchi19}.

%subsubsection {fig6}
\begin{figure}[t!]
  \centerline{\includegraphics[width=\fsize]{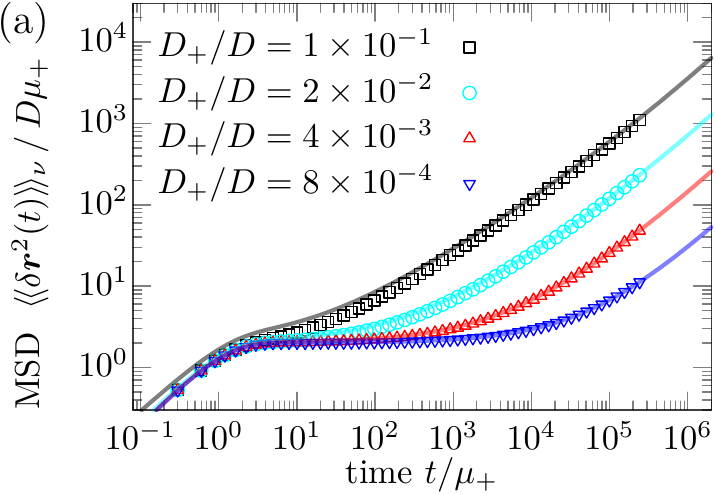}}
  \centerline{\includegraphics[width=\fsize]{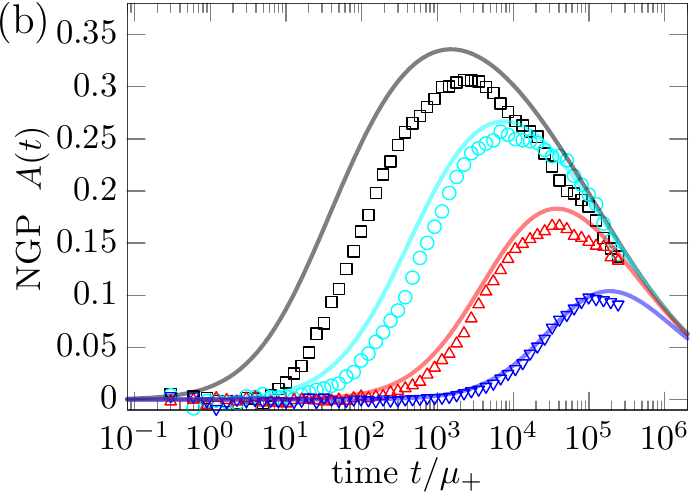}}
  \centerline{\includegraphics[width=\fsize]{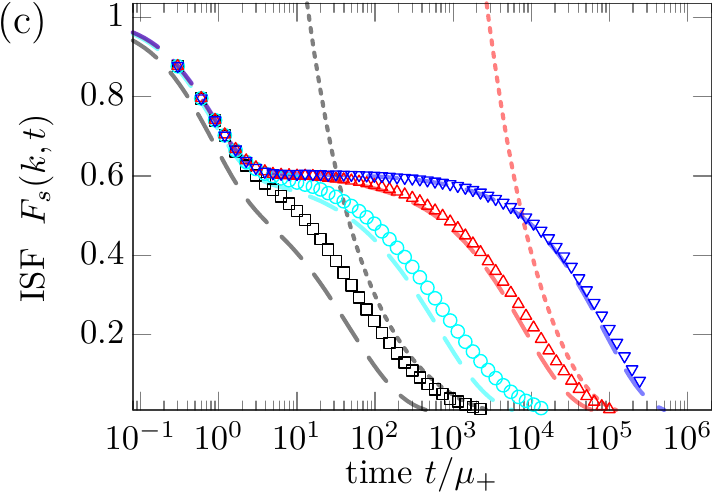}}
  \caption{\label{f.neq.alpha=075} Non-equilibrium dimer model [$\rho^{+}(\tau)$
    and $\rho^{-}(\tau)$ are an exponential distribution and a power law,
    respectively]. The symbols are data obtained by numerical integration of
    Eqs.~(\ref{e.dimer.dr(t)/dt}) and (\ref{e.dimer.dr1(t)/dt}). The fast and
    slow state diffusivities $D_\pm$ are set as
    $D_+/D = 10^{-1}, 2\times 10^{-2}, 4\times 10^{-3}, 8\times 10^{-4}$, and
    $D_- = 0.01D_+$. The power-law index $\alpha$ is fixed as $\alpha = 0.6$,
    the potential parameter $k_0$ as $\beta k_0 D\mu_+= 1$, the initial
    fractions $p_{\pm}^0$ as $p_{\pm}^0 = 0.5$, and the cutoff
    parameter $\tau_0$ as $\tau_0/ \mu_+ = 0.1$.  (a) MSD vs time. The lines are
    the theoretical prediction in Eq.~(\ref{e.emsd.noneq}).  (b) NGP vs
    time. The lines are the prediction in Eq.~(\ref{e.ngp.eq}). (c) ISF vs
    time. The lines are the prediction in Eq.~(\ref{e.F(k,t).0<alpha<1}). The
    wave number $k$ is set as $k = \sqrt{\beta k_0/2}$.  [In (b-c), results for
    the four different values of $D_+$ are displayed with the same color code as
    in (a)].  }
\end{figure}
%subsubsection {fig7}

\begin{figure}[t!]
  \centerline{\includegraphics[width=\fsize]{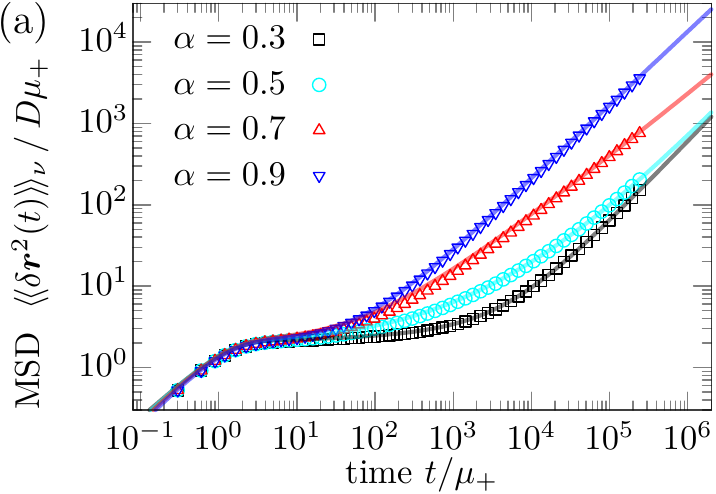}}
  \centerline{\includegraphics[width=\fsize]{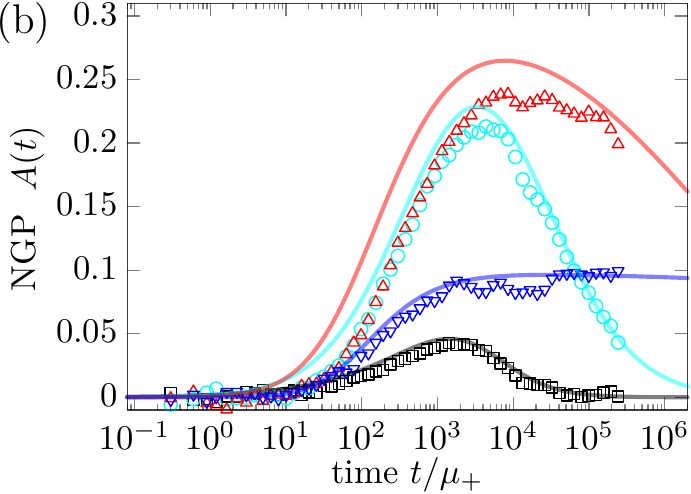}}
  \centerline{\includegraphics[width=\fsize]{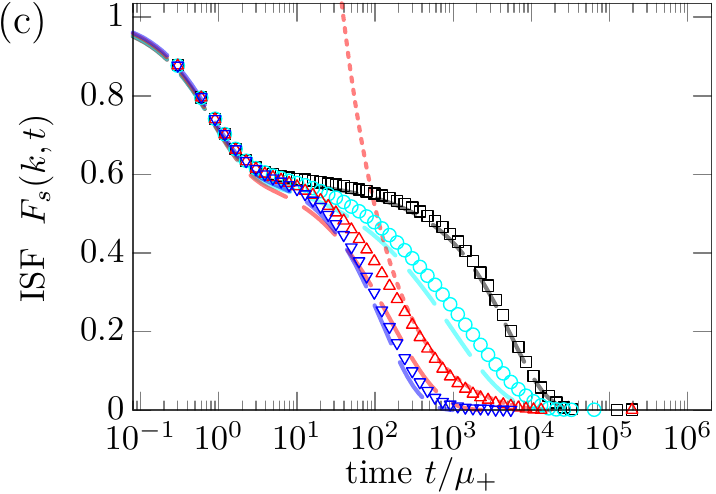}}
  
  \caption{\label{f.neq.df=003} Non-equilibrium dimer model (the same model as
    in Fig.~\ref{f.neq.alpha=075}). The power-law index is set as
    $\alpha = 0.3, 0.5, 0.7$ and $0.9$. The diffusivity $D_{\pm}$ of the two
    states are fixed as $D_+/D = 3.0 \times 10^{-2}$ and $D_- =
    0.01D_+$. Parameters $k_0$, $p_{\pm}^0$, $\tau_0$, and $k$ are the same as
    those in Fig.~\ref{f.neq.alpha=075}. (a) MSD vs time. The lines are the
    theoretical prediction in Eq.~(\ref{e.emsd.noneq}).  (b) NGP vs time. The
    lines are the prediction in Eq.~(\ref{e.ngp.noneq}). (c) ISF vs time. The
    lines are the prediction in Eq.~(\ref{e.F(k,t).0<alpha<1}). [In (b-c),
    results for the four different values of $\alpha$ are displayed with the
    same color code as in (a)].  }
\end{figure}

%subsubsection {msd}

To obtain the MSD and the NGP, we derive the ISF at small $k$.  We expand
Eq.~(\ref{e.relaxation_func.0<alpha<1.laplace}) around $k=0$, and carry out the
Laplace inversion; the resulting equation is substituted in
Eq.~(\ref{e.F(k,t)}), thereby obtaining the ISF at small $k$
{\color{black}\begin{align}
  &F_s(k, t)
  \simeq e^{-k^2 D_s(t)t}
  \notag\\[0.1cm]
  &\times\left[
  1 - k^2(A t^{\alpha} + D_-t) +  \frac {k^4(At^{\alpha}+ D_-t )^2}{2} 
  + \frac {k^4 A^2Bt^{2\alpha}}{2}
  \right]
  \notag\\[0.2cm]
  \label{e.Fs(k,t)_small_k_noneq}
  &\simeq
  \exp\left\{
  -k^2 \left[
  D_s(t) +  A t^{\alpha-1} + D_- \right] t
  + \frac {k^4 A^2B t^{2\alpha}}{2}
  \right\},
\end{align}
where we omit some lower order terms with respect to $t$; $A$ and $B$ are
constants defined as $A:= \mu_+ \Delta D_0 / a\Gamma(1+\alpha)$ and
$B=2\Gamma^2(1+\alpha)/\Gamma(1+2\alpha) - 1$.}

By using
$\left\langle\!\left\langle \delta \bm{r}^2(t) \right\rangle\!\right\rangle_\nu
= -\nabla_k^2 F_s(k,t)|_{k=0}$, we obtain the ensemble-averaged MSD as
\begin{equation}
  \label{e.emsd.noneq}
  \frac {\dbra{\delta \bm{r}^2(t)}}{2n}
  \simeq
  \left[ D_s (t) + A t^{\alpha-1} +D_- \right] t.
\end{equation}
This formula is evident again from the fact that
Eq.~(\ref{e.Fs(k,t)_small_k_noneq}) is a cumulant expansion. Alternatively, the
MSD formula can be derived with Eq.~(\ref{e.msd.app.overall.D}). In fact, from
Eqs.~(30) and (87) in Ref.~\cite{miyaguchi16}, we have
$\langle D_0(u) \rangle_\nu \simeq \alpha At^{\alpha-1} + D_-$. With this
relation and Eq.~(\ref{e.msd.app.overall.D}), we recover
Eq.~(\ref{e.emsd.noneq}).

In contrast to the equilibrium ensemble, in which only normal diffusion is
observed, the MSD for the non-equilibrium ensemble shows transient subdiffusion
at intermediate timescale. At longer timescale, normal diffusion recovers (the
long-time diffusivity is $D_-$). {\color{black}Interestingly, the same
  intermediate and long-time properties as in Eq.~(\ref{e.emsd.noneq}) are
  observed in a molecular dynamics simulation for supercooled liquids
  \cite{kob95}.} As shown in Figs.~\ref{f.neq.alpha=075}(a) and
\ref{f.neq.df=003}(a), the theoretical prediction in Eq.~(\ref{e.emsd.noneq}) is
consistent with numerical simulations, except for the case $D_+/D = 0.1$. The
slight inconsistency for $D_+/D = 0.1$ should be caused by the assumption
$D_+ \ll D$, which is used in our analysis [See the text above
Eq.~(\ref{e.f[s,s'|D].app})].

%subsubsection {tmsd}

\begin{figure}[t!]
  \centerline{\includegraphics[width=\fsize]{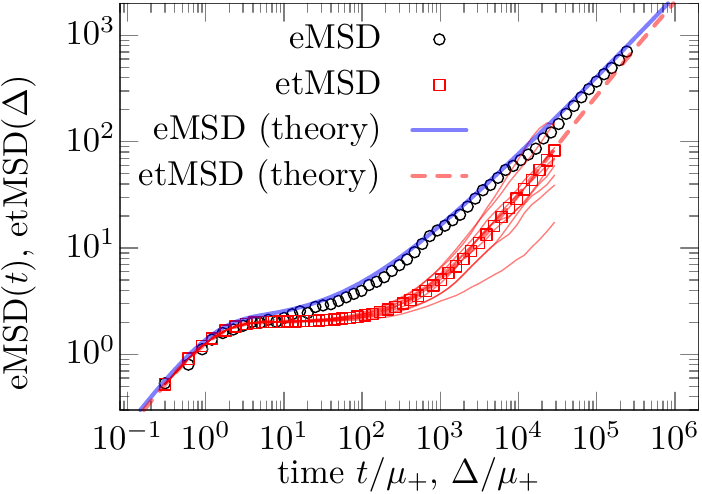}}
  
  \caption{\label{f.neq.tmsd} {\color{black}Ensemble-averaged and time-averaged
      MSDs vs time for non-equilibrium dimer model (the same model as in
      Fig.~\ref{f.neq.alpha=075}). The thick solid line and the thick dashed
      line are the theoretical predictions for the ensemble-averaged MSD
      [Eq.~(\ref{e.emsd.noneq})] and the ensemble average of time-averaged MSD
      [Eq.~(\ref{e.noneq.tmsd})], respectively. Each thin red solid line is a
      time-averaged MSD obtained from a single trajectory data; time-averaged
      MSDs for ten trajectories are displayed. The power-law index and the total
      measurement time are set as $\alpha = 0.7$ and $t=3 \times 10^5$. The
      other parameters $D_{\pm}$, $k_0$, $p_{\pm}^0$ and $\tau_0$ are the same
      as those in Fig.~\ref{f.neq.df=003}. }}
\end{figure}

{\color{black} As remarked in Sec.~\ref{s.dimer.MSD}, the ensemble-averaged and
  time-averaged MSDs do not coincide in non-equilibrium systems. Therefore, here
  a formula for the time-averaged MSD is derived by using the general expression
  in Eq.~(\ref{e.tmsd.app.overall.D}). The mean diffusivity
  $\left\langle D_0(t) \right\rangle_{\nu}$ is given by [Eqs.(30) and (87) in
  Ref.\cite{miyaguchi16}]
  \begin{equation}
    \label{e.noneq.<D0(t)>}
    \left\langle D_0(t) \right\rangle_{\nu}
    \approx
    D_- + \Delta D_0 \frac {\mu_+ t^{\alpha-1}}{a \Gamma(\alpha)}.
  \end{equation}
  Thus, $D_0(t)$ is a non-stationary process; the origin of the non-stationarity
  can be understood from the fact that the diffusive state is increasingly stuck
  in the slow state $D_-$ as $t$ increases \cite{korabel10b, miyaguchi16}.
  
  Inserting Eq.~(\ref{e.noneq.<D0(t)>}) into Eq.~(\ref{e.tmsd.<D0(u;t)>_nu}), we
  obtain an explicit formula for
  $\left\langle \overline{D}_0(u; t) \right\rangle_{\nu}$. Then, substituting
  this formula into Eq.~(\ref{e.tmsd.app.overall.D}), we have the ensemble
  average of the time-averaged MSD as
  \begin{equation}
    \label{e.noneq.tmsd}
    \frac {\dbra{\overline{\delta \bm{r}}^2(\Delta; t)}}{2n}
    \simeq
    \left[D_s(\Delta) + A t^{\alpha-1} + D_- \right] \Delta,
  \end{equation}
  where $\Delta \ll t$ is assumed. Thus, the time-averaged MSD has a form
  different from the ensemble-averaged one [Eq.~(\ref{e.emsd.noneq})]; this is a
  natural consequence of the non-stationarity of $D_0(t)$
  [Eq.~(\ref{e.noneq.<D0(t)>})].

  In particular, the time-averaged MSD does not exhibit subdiffusion in contrast
  to the ensemble-averaged MSD. It is also prominent that the time-averaged MSD
  depends on the measurement time $t$; such a property is referred to as an
  aging effect \cite{schulz14}. This aging effect occurs at an intermediate
  timescale (with respect to the measurement time $t$), at which the
  ensemble-averaged MSD shows the subdiffusion.

  In Fig.~\ref{f.neq.tmsd}, time dependences of the ensemble-averaged MSD
  [Eq.~(\ref{e.emsd.noneq})] and the ensemble average of the time-averaged MSD
  [Eq.~(\ref{e.noneq.tmsd})] are displayed with thick solid and dashed lines.
  It is clear that these two MSDs behave differently at the intermediate
  timescale. In contrast, at short and long times, these two MSDs coincide.

  Moreover, in Fig.~\ref{f.neq.tmsd}, numerical results of the time-averaged
  MSDs are also presented with thin red lines; each red line is calculated from
  a single trajectory data. It is clear that the time-averaged MSDs show
  trajectory-to-trajectory fluctuations. Relaxation properties of these
  fluctuations should be closely related to non-Gaussianity of displacement
  distributions \cite{miyaguchi16}.}

%subsubsection {fig8}

%subsubsection {ngp}

In the same way as the ensemble-averaged MSD, the NGP [Eq.~(\ref{e.ngp-def})]
can be derived with
$\dbra{\delta \bm{r}^4} = \nabla_k^2\nabla_k^2 F_s(k,t)|_{k=0}$ as
\begin{equation}
  \label{e.ngp.noneq}
  A(t) \simeq 
  \frac {A^2Bt^{2\alpha-2}}{\left(1/\beta k_0 t + D_- + At^{\alpha-1}\right)^2},
  %% \left[\frac {2\Gamma^2(1+\alpha)}{\Gamma(1+2\alpha)} - 1\right]
  %% \frac {A^2t^{2\alpha-2}}{\left(1/\beta k_0 t + D_- + At^{\alpha-1}\right)^2},
\end{equation}
where the short-time diffusivity $D_s(t)$ is replaced with its asymptotic form
at long times $D_s(t) \simeq 1/\beta k_0t$, because
Eq.~(\ref{e.relaxation_func.0<alpha<1.laplace}) is valid only for large $t$
(i.e., $t \gg \mu_\pm, \,a^{1/\alpha}$). This function shows a unimodal shape as
shown in Figs.~\ref{f.neq.alpha=075}(b) and \ref{f.neq.df=003}(b), in which
Eq.~(\ref{e.ngp.noneq}) is compared with the simulations. The agreement is
fairly good, but there are some {\color{black}deviations}, which might be due to
lower order corrections neglected in deriving
Eq.~(\ref{e.Fs(k,t)_small_k_noneq}).

%subsubsection {isf}

Next, let us derive a formula for the ISF $F_s(k,t)$ without the assumption
of $k$ being small.  With the Laplace inversion of the first and second terms of
Eq.~(\ref{e.relaxation_func.0<alpha<1.laplace}) (the third and fourth terms are
neglected, because they are lower order corrections in terms of $t$), the
relaxation function is given by [Eqs.~(63) and (65) in Ref.~\cite{miyaguchi19}]
\begin{align}
  \label{e.relaxation_func.0<alpha<1}
  \left\langle  e^{-k^2 \int^t_0 D_0(u)du} \right\rangle_\nu
  \simeq
  \begin{cases}
    e^{-k^2 \left[D_- +  A t^{\alpha-1}\right]t},
    & (t \sim \frac {1}{D_+k^2}),\\[0.1cm]
    \frac {\Delta D_0}{k^2D_+^{2}} \frac {a}{\mu_+}
    \frac {e^{-k^2 D_- t}}{\Gamma(1-\alpha) t^{\alpha}},
    & (t \gg \frac {1}{D_+k^2}).
  \end{cases}
\end{align}
By using Eqs.~(\ref{e.F(k,t)}) and (\ref{e.relaxation_func.0<alpha<1}), the ISF
is obtained as
\begin{align}
  \label{e.F(k,t).0<alpha<1}
  F_s(k, t)
  \simeq
  \begin{cases}
    e^{-k^2 \left[D_s(t) + D_- + A t^{\alpha-1}\right]t},
    & (t \sim \frac {1}{D_+k^2}),\\[0.1cm]
    \frac {\Delta D_0}{k^2D_+^{2}} \frac {a}{\mu_+}
    \frac {e^{-k^2 [D_s(t) + D_- ]t}}{\Gamma(1-\alpha) t^{\alpha}},
    & (t \gg \frac {1}{D_+k^2}),
  \end{cases}
\end{align}
which shows an exponential relaxation at short time, a stretched-exponential
relaxation at intermediate time, and a power-law relaxation at long
time.

As in the equilibrium case, the first equation in Eq.~(\ref{e.F(k,t).0<alpha<1})
exhibits a two-step relaxation as shown in Figs.~\ref{f.neq.alpha=075}(c) and
\ref{f.neq.df=003}(c). The theoretical prediction (the dashed lines) is
consistent with the numerical simulation except at long time, at which the
second equation in Eq~.~(\ref{e.F(k,t).0<alpha<1}) shows good agreement [the
second equation is shown by dotted lines only for the cases $D_+/D =10^{-1}$ and
$4 \times 10^{-3}$ in Fig.~\ref{f.neq.alpha=075}(c) and $\alpha=0.7$ in
Fig.~\ref{f.neq.df=003}(c)].

\section {Discussion}\label{s.discussion}

Several models with both the memory effect and the fluctuating diffusivity have
been proposed and studied so far \cite{slezak18, wang20, wang20b, sabri20,
  janczura21, dieball22, goswami22}, but physical backgrounds of these models,
such as the fluctuation-dissipation relation, have been unclear. In this paper,
we proposed the GLEFD [Eq.~(\ref{e.glefd})], and showed that it satisfies the
generalized fluctuation-dissipation relation [Eq.~(\ref{e.phi(t,t')})]. By
utilizing the Markovian embedding method, a numerical integration scheme of the
GLEFD was also presented. With the physical background, the GLEFD
[Eqs.~(\ref{e.glefd}) and (\ref{e.phi(t,t')})] should become an important model
to explain complex diffusion processes observed in single-particle-tracking
experiments and molecular dynamics simulations.

To explain the non-Gaussianity in the viscoelastic dynamics, a random potential
energy (quenched disorder) has also been frequently employed \cite{goychuk20,
  goychuk21}. It is probable that such an approach is related to the one
presented in this article, because quenched disorder can be approximated with
annealed disorder for some systems with dimension higher than two
\cite{machta85, bouchaud92}, and the annealed disorder might be modeled by the
fluctuating diffusivity \cite{uneyama15}. It is also worth noticing that
annealed models are easier to handle theoretically than quenched models
\cite{bouchaud90}.

Moreover, as special cases of the GLEFD, the FBMFD and the dimer model were
investigated in detail. The FBMFD, which shows the subdiffusion, non-Gaussianity
and stretched-exponential relaxation, must be an important model in biological
applications. This is because subdiffusion and non-Gaussianity are observed in
many single-particle-tracking experiments inside cytoplasm \cite{he08, parry14,
  lampo17, sabri20, janczura21} and molecular dynamics simulations
\cite{akimoto11, yamamoto14, jeon16, yamamoto17}; importantly, in some
experiments and simulations, the subdiffusion is attributed to the FBM-like
anti-persistent correlation \cite{magdziarz09, thapa18, lampo17, akimoto11,
  jeon16}. The origin of the non-Gaussianity is still controversial, but, in our
approach, it is attributed to the fluctuating diffusivity.

In the FBMFD in this article, the power-law index $\alpha$ is constant, whereas
some experiments have revealed that this parameter is also random \cite{sabri20,
  janczura21}. To incorporate a time-dependent fluctuation in $\alpha$ in our
approach, $k_i$ in Eq.~(\ref{e.gle.wo.fd.markov.2}) should be fluctuating and
depend on time. This however causes a violation of the detailed balance, because
the potential energy changes with time. Thus, it is an interesting question
whether equilibrium dynamics can or cannot explain the randomness in the
power-law index $\alpha$.

The dimer model is one of the simplest model of the GLEFD, and analytically
tractable to some extent. In fact, only this case can be described by a single
equation without an integral term [Eq.~(\ref{e.gle})], with which we can see a
complicated interplay between the memory effect and the fluctuating diffusivity
[Eq.~(\ref{e.<eta(t)eta(t')>})].  Formulas for the MSD, the NGP, and the ISF in
the dimer model were also derived analytically, and it was found that they are
fairly consistent with numerical simulations. In particular, for the
non-equilibrium ensemble, the dimer model exhibits subdiffusion as well as 
stretched-exponential relaxation at intermediate timescales.

Both the FBMFD and the non-equilibrium dimer model show subdiffusion,
non-Gaussianity (with a unimodal shape in the NGP), and stretched-exponential
relaxation. In addition to these features, the dimer model also exhibits
plateaus in the MSD and the ISF; these features are similar to those observed in
supercooled liquids and glassy systems \cite{gotze_book}. However, it is
numerically reported in Ref.~\cite{kob95, kob95b} that these properties are
observed even in equilibrium systems. In contrast, for the dimer model,
subdiffusion and stretched-exponential relaxation are observed only for the
non-equilibrium initial conditions, and thus the mechanism giving rise to the
above properties in the dimer model might well be different from that in the
supercooled liquids and glassy systems.
%% (a trapping mechanism similar to that in the continuous-time random walk
%% \cite{uneyama15})

Note however that we only studied the fast switching limit for the dimer model
with power-law sojourn times. For the case in which timescales of switching and
diffusion are comparable, analytical results are limited to the equilibrium
dimer model in which the sojourn-time distributions $\rho^{\pm}(\tau)$ are both
given by exponential distributions [Fig.~\ref{f.exp-dist}]. Thus, an interesting
question for future studies would be to clarify the properties without the fast
switching assumption. In addition, throughout the analysis of the dimer model
[Secs.~\ref{s.dimer} and \ref{s.two-state-model}], we assume $D_0(t) \ll D$; it
should be also important to study the dimer model without this assumption.

To incorporate the fluctuating diffusivity into the GLE, we used the Markovian
embedding in this paper, but there should be other ways to do so. For example,
it is possible to incorporate it by assuming that the memory is reset at each
switching event.  There would be some situations in which resetting the memory
is a natural choice \cite{dieball22}, though evolution equations without
resetting (such as the GLEFD in this article) might be more natural in other
situations; an example is monomer dynamics of a flexible polymer, in which the
memory effect originates from internal degrees of freedom of the polymer
\cite{panja10b}, and the fluctuating diffusivity originates from hydrodynamic
interactions \cite{miyaguchi17, yamamoto21} or polymer size fluctuations
\cite{nampoothiri21, nampoothiri22}.

But, whether it is possible to derive the GLEFD from microscopic models such as
polymer systems should be explored in future studies. {\color{black}Because the
  GLE is derived from microscopic equations of motion with the projection
  operator method \cite{gotze_book, hansen90}, and thus the auxiliary variables
  should be related to the fast variables which are projected out. For example,
  the middle monomer of the Rouse polymer can be described by
  Eqs.~(\ref{e.gle.wo.fd.markov.1}) and (\ref{e.gle.wo.fd.markov.2}), where the
  auxiliary variables $\bm{r}_i(t)$ are normal modes of the Rouse polymer
  \cite{panja10b}. If the hydrodynamic interactions between the monomers are
  further taken into account, the diffusivities of the normal modes might be
  fluctuating \cite{miyaguchi17}. These points will be investigated in more
  detail and reported elsewhere.}

Although the GLEFD is constructed in the presence of the external force
$\bm{F}_e$ [Eq.~(\ref{e.glefd})], the FBMFD and the dimer model were studied in
the absence of $\bm{F}_e$. Thus, effects of external forces such as a constant
force and a force due to a harmonic potential should be explored in future
work. Moreover, beyond the analysis of the NGP, elucidation of displacement
distributions for the GLEFD is also an important future subject
\cite{lanoislee18, barkai20, soria21}. {\color{black}A quantity called
  codifference may also be useful to characterize the non-Gaussianity
  \cite{slezak19}.}

{\color{black} Fluctuation properties of the time-averaged MSD are not elucidated
  in this work, but such properties are quite important in that they are closely
  related to the non-Gaussianity of the displacement, and that fluctuation
  analysis of the time-averaged MSD would be more efficient than the NGP method
  \cite{miyaguchi17, yamamoto21}. Fluctuation properties of the time-averaged
  MSD have been studied with its relative standard deviation or relative
  variance (the latter is referred to as an ergodicity breaking parameter
  \cite{metzler14}). For normal diffusion with fluctuating diffusivity, such
  parameters have been shown to behave similarly to the NGP $A(t)$ at long times
  \cite{miyaguchi16}, but further studies are necessary for subdiffusive
  systems.}

In this article, the fluctuating diffusivity is assumed to be a scalar.  But,
many diffusion processes are described by diffusivity given by a second rank
tensor \cite{dhont96, miyaguchi17}. Therefore, a generalization to the tensor
diffusivity should be studied in future. {\color{black}Moreover, in all the
  simulations in this article, the fluctuating diffusivity is assumed to be
  two-state processes. Instead of these two-state processes, it is interesting
  to use a process of which stationary distribution is consistent with
  experimentally observed distributions such as the exponential distribution
  \cite{lampo17, chechkin17, lanoislee18}.}

{\color{black} In GLEFD [Eq.~(\ref{e.glefd})], the fluctuating diffusivity is
  incorporated only into the correlated noise $\bm{\xi}_c(t)$ and the white
  Gaussian noise $\bm{\xi}(t)$ is left unchanged. But, it is also interesting to
  study the opposite case in which the diffusivity due to the white noise
  $\bm{\xi}(t)$ is fluctuating and that due to the correlated noise
  $\bm{\xi}_c(t)$ is not fluctuating. Such a model might be relevant for
  tagged-particle motion in single-file diffusion \cite{kollmann03} with wall
  effects \cite{alexandre22}.}  Furthermore, we studied the GLEFD at the
overdamped limit, but it is also possible to incorporate the fluctuating
diffusivity into the underdamped GLE \cite{goychuk09}.

The procedure used to obtain the FBMFD is readily generalized into an arbitrary
memory kernel. In fact, if an even function $\phi(t)$ is a memory kernel of a
GLE without the fluctuating diffusivity and the property in
Eq.~(\ref{e.glefd.two-state}) is assumed, then the memory kernel $\phi(t, t')$
of the corresponding GLEFD is given by
\begin{equation}
  \label{e.phi(t,t').general}
  \phi(t, t')
  =
  \phi\left(\int_{t'}^{t}\kappa(u)du\right).
\end{equation}
Note that Eqs.~(\ref{e.phi(t).power.markov}) and
(\ref{e.phi(t,t').two-state.power.approx}) satisfy the above relation.

Unfortunately, it is difficult at present to theoretically analyze the GLEFD,
because the integral term of the GLEFD [Eq.~(\ref{e.glefd})] is not of a
convolution form and consequently the Laplace transform cannot be utilized
{\color{black}[Note also that, in spite of this non-convolution form, the
  time-translation symmetry is not broken, if the system is in equilibrium. See
  Eq.~(\ref{e.<phi(t,t')>_nu})]}. Hence, developing theoretical tools,
{\color{black}which includes an approach through Fokker--Planck-like equations
  \cite{fox77},} to analyze the GLEFD should be one of the most important future
tasks.

{\color{black}It is interesting to examine whether the generalized
  fluctuation-dissipation relation in Eq.~(\ref{e.phi(t,t')}) is valid in more
  general diffusivity fluctuations. For example, suppose that the correlated
  noise $\bm{\xi}_c(t)$ in the GLEFD [Eq.~(\ref{e.glefd})] is given by
  \begin{equation}
    \label{e.multiplicative}
    \bm{\xi}_c(t) = B(t) \bm{\xi}_c^0(t),
  \end{equation}
  where $B(t)$ is a stochastic process, and $\bm{\xi}_c^0(t)$ is a correlated
  Gaussian noise [a similar system was studied in Refs.~\cite{wang20, wang20b}
  without the integral term and the white noise $\bm{\xi}(t)$]. Then, it is
  tempting to conjecture from Eq.~(\ref{e.phi(t,t')}) that the
  fluctuation-dissipation relation would be given by
  \begin{equation}
    \label{e.multiplicative.phi}
    \phi(t,t') \bm{I}
    =
    \left\langle \bm{\xi}_c(t)\bm{\xi}_c(t') \right\rangle
    =
    B(t)B(t')\left\langle \bm{\xi}_c^0(t)\bm{\xi}_c^0(t') \right\rangle.
  \end{equation}

  It is possible to show that the conjecture is true with a Markovian embedding
  method. In fact, the Markovian embedding in Sec.~\ref{s.markov-embed} can be
  utilized again except Eq.~(\ref{e.(ri-r)}), which is replaced with
  \begin{equation}
    \label{e.multiplicative.(ri-r)}
    k_i'(\bm{r}_i - \bm{r})
    =
    \sqrt{D}\bm{\xi}_i - k_i' \int_0^t e^{-\nu_i (t-t')} B(t')\dot{\bm{r}}(t') dt'.
  \end{equation}
  Then, Markovian equations of motion are obtained as
  \begin{equation}
    \label{e.multiplicative.dR/dt}
    \frac {d\bm{R}(t)}{dt} = \beta\bm{L}\cdot\bm{F} + \bm{\Xi}, 
  \end{equation}
  where a supervector notation
  $\bm{R}(t):= (\bm{r}, \bm{r}_0,\dots, \bm{r}_{N-1})$ is employed

  In Eq.~(\ref{e.multiplicative.dR/dt}), $\beta\bm{L}$ is a mobility tensor
  defined by
  \begin{align}
    \label{e.multiplicative.L}
    \bm{L} =
    \begin{pmatrix}
      D \bm{I}  & DC \bm{I}         & \dots & DC^2 \bm{I}          \\
      DC \bm{I} & (D_0+DC^2) \bm{I} & \dots & DC^2 \bm{I}          \\
      \vdots    &                   &       &                      \\
      DC \bm{I} & DC^2 \bm{I}       & \dots & (D_{N-1}+DC^2) \bm{I} \\
    \end{pmatrix},
  \end{align}
  where $C(t):= 1-B(t)$. Moreover, $\bm{F}$ is a (non-random) force vector given
  by $\bm{F} := (\bm{F}_t+\bm{F}_e, \bm{F}_0, \dots, \bm{F}_{N-1})$. Here,
  $\bm{F}_i$ are harmonic forces exerted on the auxiliary variables $\bm{r}_i$
  and are defined by $\bm{F}_i := k_i(\bm{r}- \bm{r}_i)$; and $\bm{F}_t$ is a
  resultant force exerted on the original variable $\bm{r}$ and is defined by
  $\bm{F}_t := - \sum_{i=0}^{N-1} \bm{F}_i$. Finally, $\bm{\Xi}$ is a noise
  vector defined by
  \begin{align}
    \bm{\Xi} = 
    \begin{pmatrix}
      \sqrt{2D}\bm{\xi}                                       \\
      \sqrt{2D}C \bm{\xi} + \sqrt{2D_0} \bm{\eta}_{0}         \\
      \vdots                                                  \\
      \sqrt{2D}C \bm{\xi} + \sqrt{2D_{N-1}} \bm{\eta}_{N-1} \\
    \end{pmatrix},
  \end{align}
  where $\bm{\eta}_i(t)$ is the white Gaussian noise defined in
  Eq.~(\ref{e.<eta_i(t)eta_j(t')>}). If $C(t)\equiv 0$,
  Eq.~(\ref{e.multiplicative.dR/dt}) is equivalent to
  Eqs.~(\ref{e.gle.wo.fd.markov.1}) and (\ref{e.gle.wo.fd.markov.2}).

  Then, it is readily shown that Eq.~(\ref{e.multiplicative.dR/dt}) satisfies
  the fluctuation-dissipation relation
  $\left\langle \bm{\Xi}(t) \bm{\Xi}(t') \right\rangle = 2 \bm{L} \delta(t-t')$
  \cite{doi86}. Thus, a generality of Eq.~(\ref{e.phi(t,t')}) should be
  promising, but further studies are obviously required. Moreover, it should be
  interesting per se to study the GLEFD [Eq.~(\ref{e.glefd})] with
  Eqs.~(\ref{e.multiplicative}) and (\ref{e.multiplicative.phi}). Note that
  Eq.~(\ref{e.multiplicative.dR/dt}) can be utilized as a scheme for numerical
  integration of this system. }

%section {acknowledgments}
\begin{acknowledgments}
  The author wishes to acknowledge Dr. Takuma Akimoto for valuable
  comments. This work was supported by Grant-in-Aid (KAKENHI) for Scientific
  Research C (Grant No. JP18K03417 and JP22K03436). 
\end{acknowledgments}

\appendix {}
\section {Fractional Brownian motion}\label{app.fbm}

In this Appendix, the FBM without the fluctuating diffusivity
[Eqs.~(\ref{e.gle.wo.Fe}) and (\ref{e.phi(t).power})] is briefly reviewed.
Carrying out the double Laplace transforms ($t \leftrightarrow s$ and
$t' \leftrightarrow s'$) of Eqs.~(\ref{e.<xi-xi>.simple}) and
(\ref{e.<xi_c-xi_c>.simple}), we have fluctuation-dissipation relations in the
Laplace domain \cite{pottier03}
\begin{align}
  \label{e.<xi-xi>.simple.laplace}
  \Bigl\langle \hat{\bm{\xi}}(s) \hat{\bm{\xi}}(s') \Bigr\rangle
  &=
  \frac {1}{s + s'} \bm{I},
  \\[0.1cm]
  \label{e.<xi_c-xi_c>.laplace}
  \left\langle \hat{\bm{\xi}}_c^0(s) \hat{\bm{\xi}}_c^0(s') \right\rangle
  &=
  \frac {\hat{\phi} (s) + \hat{\phi} (s')}{s+s'} \bm{I}.
\end{align}
Moreover, the Laplace transform of Eq.~(\ref{e.gle.wo.Fe}) is given by
\begin{equation}
  \label{e.app.r(s)}
  \hat{\bm{r}}(s) =
  \sqrt{D}
  \frac {\sqrt{2} \hat{\bm{\xi}}(s) + \hat{\bm{\xi}}_c(s)}
  {s [1+ \hat{\phi}(s)]}
\end{equation}
where $\bm{r}(0)= 0$ is assumed, and consequently $\delta \bm{r}(t) = \bm{r}(t)$
in the following.  With these relations and the independence of $\hat{\bm{\xi}}$
and $\hat{\bm{\xi}}_c^0$, we obtain
\begin{equation}
  \label{e.app.<r(s)r(s')>}
  \left\langle \hat{\bm{r}}(s)\hat{\bm{r}}(s') \right\rangle
  =
  D\frac {\frac {1}{1+ \hat{\phi}(s)} + \frac {1}{1+ \hat{\phi}(s')}}
  {ss'(s+s')}\bm{I},
\end{equation}

If the memory kernel $\phi(t)$ is the power-law form in
Eq.~(\ref{e.phi(t).power}), its Laplace transform is given by
$\hat{\phi}(s) = A \Gamma (1-\alpha) s^{\alpha-1}$\,$(0<\alpha<1)$. Using this
relation and Eq.~(\ref{e.app.<r(s)r(s')>}), we obtain
\begin{align}
  \left\langle \hat{\bm{r}}(s)\hat{\bm{r}}(s') \right\rangle
  \simeq
  \begin{cases}
    \frac {2D}{ss'(s+s')}\bm{I}, & [1 \ll \hat{\phi}(s), \hat{\phi}(s')],
    \\[0.1cm]
    \frac {D_{\alpha} \Gamma(1+\alpha) (s^{1-\alpha} + s'^{1-\alpha})}
    {ss'(s+s')} \bm{I},
    & [1 \gg \hat{\phi}(s), \hat{\phi}(s')].
  \end{cases}
\end{align}
Then, the double Laplace inversions give the following asymptotic formulas at
short and long times:
\begin{align}
  \left\langle \bm{r}(t) \bm{r}(t') \right\rangle
  \simeq
  \begin{cases}
    {2D} \mathrm{min}(t,t')\bm{I}, & (t, t' \ll t_c),
    \\[0.1cm]
    D_{\alpha} [t^{\alpha} + t'^{\alpha} - |t-t'|^{\alpha}]\bm{I},
    & (t, t' \gg t_c),
  \end{cases}
\end{align}
where $t_c:= [\Gamma(1+\alpha) D_{\alpha}/D]^{1/(1-\alpha)}$ is the crossover
time. By putting $t=t'$ and taking contractions of the tensors, we obtain the
MSD in Eq.~(\ref{e.gle.msd}).

Likewise, for the case in which $t'\ll t_c \ll t$, we have
$\langle \bm{r}(t) \bm{r}(t') \rangle \simeq D_\alpha[t^{\alpha} -
(t-t')^{\alpha}]$. Autocorrelation of the velocity $\bm{v}(t) := d \bm{r}(t)/dt$
is obtained by differentiating the above equation with $t$ and $t'$ and then
putting $t'=0$ as
$\langle \bm{v}(t) \bm{v}(0) \rangle \simeq \alpha(\alpha-1)t^{\alpha-2} \bm{I}$
for $t \gg t_c$. Thus, the velocity shows negative (anti-persistent) correlation
at long times.

\section {Exact MSD formula for dimer model}
\label{s.exact-msd}

For the dimer model, a formula for the MSD
[Eq.~(\ref{e.msd.app.overall.D.steady})] is derived under the assumption
$D_0(t) \ll D$ for any $t$. But, if $D_0(t)$ is a stationary process, it is also
possible to derive an exact formula as shown below.  Let us study the two-state
model in which the sojourn-time distributions $\rho^{\pm}(\tau)$ are exponential
distributions. From Eqs.~(\ref{e.f(s)}) and (\ref{e.relaxation_func.exp-dist}),
we have
\begin{equation}
  \label{e.msd.non-markov.exp-dist}
  \frac {f(t)}{D \beta k_0}
  =
  A_+ e^{-(\beta k_0 D + \lambda_+)t}
  +
  A_- e^{-(\beta k_0 D + \lambda_-)t},
\end{equation}
where $k^2$ in $\lambda_{\pm}$ [Eq.~(\ref{e.dimer.exp-dist.lambda})] and
$A_{\pm}$ [Eq.~(\ref{e.dimer.exp-dist.A})] should be replaced with $\beta k_0$.

Then, the MSD can be calculated with Eqs.~(\ref{e.emsd.stationary}) and
(\ref{e.msd.non-markov.exp-dist}) as
\begin{equation}
  \label{e.msd.non-markov.exp-dist.f}
  \frac {\dbra{\delta \bm{r}^2(t)}}{2n}
  =
  D
  \left[
  t  - \frac {\beta k_0D}{2} t^2 \sum_{l = \pm} A_lg_d ((\beta k_0D+\lambda_{l})t)
  \right],
\end{equation}
where $g_d(t)$ is the Debye function defined in Eq.~(\ref{e.debye-func}).  The
equation (\ref{e.msd.non-markov.exp-dist.f}) has the same asymptotic forms as
those of Eq.~(\ref{e.msd.app.overall.D.steady}). Note also that
Eq.~(\ref{e.msd.app.overall.D.steady}) is valid if $D_{\pm} \ll D$, whereas
Eq.~(\ref{e.msd.non-markov.exp-dist.f}) does not need such a restriction. For
cases in which there are more relaxation modes (See Ref.~\cite{uneyama19} for an
example), a result similar to Eq.~(\ref{e.msd.non-markov.exp-dist.f}) can be
readily obtained.

\section {Sojourn time distribution}\label{app.sojourn-time-pdf}
%subsection {equilibrium sojourn-time distribution}
In the FBMFD and the dimer model, the fluctuating diffusivity $\kappa(t)$
[Eq.~(\ref{e.glefd.two-state})] and $D_0(t)$ [Eq.~(\ref{e.D0(t).two-state})] are
defined as two-state processes. Switching between the two states can be
characterized by the sojourn-time distributions of the two states $\rho^{\pm}(\tau)$. In
the case studies in this article, we assume that $\rho^{\pm}(\tau)$ follow
either exponential distributions or power-law distributions. Thus, in the
following, we briefly summarize switching processes and these distributions (See
Ref.~\cite{miyaguchi19} for detail).

We suppose that observation of the two-state process starts at $t=0$, but if the
process is in equilibrium at $t=0$, the process should have started at
$t=-\infty$. Therefore, $t=0$ is not a renewal time (the renewal time is the
time at which the state just switches from one state to the other). In other
words, the first sojourn time does not follow the sojourn-time distribution
$\rho^{\pm}(\tau)$. Instead of $\rho^{\pm}(\tau)$, the first sojourn time
follows the equilibrium distribution $\rho^{\pm, \mathrm{eq}}(\tau)$, which is
defined through its Laplace transform as (See Ref.~\cite{miyaguchi19} for a
derivation)
\begin{equation}
  \label{e.app.rho^eq}
  \hat{\rho}^{\pm, \mathrm{eq}}(s)
  =
  \frac {1 - \hat{\rho}^{\pm}(s)}{\mu_{\pm}s}.
\end{equation}
Note that the equilibrium distribution $\rho^{\pm, \mathrm{eq}}(\tau) $ exists only if
the mean sojourn time $\mu_{\pm}$ is finite.

\subsection {Exponential distribution}

%% We assume that the sojourn-time distribution of the fast state $\rho^+ (\tau)$ follows
%% the exponential distribution.
As sojourn-time distributions $\rho^{\pm}(\tau)$ for the FBMFD and the dimer model,
we employ the exponential distribution
\begin{equation}
  \label{e.exp-dist}
  \rho^{\pm} (\tau)
  =
  \frac {1}{\mu_{\pm}} \exp\left(- \frac {\tau}{\mu_{\pm}}\right).
  %% =
  %% k_+ \exp\left(- k_+{\tau}\right).
\end{equation}
The Laplace transform of $\rho^{\pm} (\tau)$ is given by
($\tau \leftrightarrow s$)
\begin{equation}
  \label{e.laplace.exp-dist}
  \hat{\rho}^\pm (s)
  = \frac {1}{1 + \mu_\pm s}.
  %  = \frac {k_+}{k_+ + s}.
\end{equation}
For the exponential distribution, the equilibrium distribution
$\rho^{\pm, \mathrm{eq}} (\tau)$ is equivalent to $\rho^{\pm} (\tau)$
\cite{miyaguchi19}, because
\begin{equation}
  \label{e.laplace.exp-dist.stationality}
  \hat{\rho}^{\pm, \mathrm{eq}} (s)
  =
  \frac {1 - \hat{\rho}^{\pm}(s)}{\mu_{\pm}s}
  =
  \hat{\rho}^{\pm} (s),
\end{equation}
where Eq.~(\ref{e.app.rho^eq}) is used. As a result, the first sojourn time
$\tau_1$ follows the same distribution [Eq.~(\ref{e.exp-dist})] as that of the
later sojourn times of the same state $\tau_k\, (k=3, 5, \dots)$.

\subsection {Power law distribution}

As a sojourn-time distribution of the slow state $\rho^-(\tau)$ in the dimer model, the
power-law distribution is also employed. The power-law distribution is defined
by
\begin{equation}
  \label{e.rho(t)}
  \rho^- (\tau)
  \underset{\tau\to\infty}{\simeq}
  \frac {a}{|\Gamma (-\alpha)|\tau^{1+\alpha}}, 
\end{equation}
%% where double-sign corresponds (In all equations in what follows, double-signs
%% correspond).
%where $c_{\pm}$ is defined by $c_{\pm} = a_{\pm} / |\Gamma (-\alpha_{\pm})|$,
where $\alpha$ is the power-law index with $0<\alpha<2$, $a$ is a scale factor,
and $\Gamma (-\alpha)$ is the Gamma function. Asymptotic forms of its Laplace
transform $\hat{\rho}^- (s) := \int_0^{\infty} d\tau\, e^{-\tau s} \rho^-(\tau)$
at small $s$ are given by
\begin{alignat}{2}
  \label{e.rho(s).asymptotic.alpha<1}
  \hat{\rho}^-(s)
  & \underset{s \to 0}{\simeq} 1 - a s^{\alpha} + o(s^{\alpha}), \quad
  & (0 < \alpha < 1),
 \\[0.1cm]
  \label{e.rho(s).asymptotic.alpha>1}
  \hat{\rho}^-(s)
  & \underset{s \to 0}{\simeq}  1 - \mu_- s + a s^{\alpha}  + o(s^{\alpha}),\quad
  & (1 < \alpha < 2),
\end{alignat}
where $o(s^{\alpha})$ is the Landau notation.

For $0 < \alpha < 1$, the mean sojourn time $\mu_-$ diverges, and therefore the
equilibrium distribution $\rho^{-, \mathrm{eq}} (\tau)$ does not exist. In
contrast, for $1 < \alpha < 2$, $\mu_-$ is finite, and consequently the
equilibrium distribution $\rho^{-, \mathrm{eq}} (\tau)$ exists; in fact, from
Eqs.~(\ref{e.app.rho^eq}) and (\ref{e.rho(s).asymptotic.alpha>1}), it is given
by \cite{miyaguchi19}
\begin{equation}
  \label{e.equilibrium.power-law}
  \hat{\rho}^{-, \mathrm{eq}} (s)
  =
  \frac {1 - \hat{\rho}^{-}(s)}{\mu_-s}
  \underset{s \to 0}{\simeq} 
  1
  -
  \frac {a}{\mu_-} s^{\alpha-1}.
\end{equation}

In numerical simulations, we employ the following power-law distribution for
$0 < \alpha <2$
\begin{equation}
  \label{e.simulation.rho(tau)}
  \rho^- (\tau) = \frac {\alpha \tau_0^{\alpha}}{\tau^{1+\alpha}},
  \qquad
  (\tau_0 \leq \tau < \infty)
\end{equation}
where $\tau_0$ is a cutoff time for short sojourn times. By comparing
Eq.~(\ref{e.simulation.rho(tau)}) with a general form in Eq.~(\ref{e.rho(t)}),
it is found that $a$ is given by $a= \tau_0^{\alpha}|\Gamma(1-\alpha)|$. For
$1< \alpha < 2$, the mean sojourn time $\mu_-$ of this distribution
[Eq.~(\ref{e.simulation.rho(tau)})] is given by
$\mu_- = \alpha \tau_0 / (\alpha - 1)$. Sojourn times $\tau_k$ which follow the
equilibrium distribution $\rho^{\pm, \mathrm{eq}}(\tau)$ are generated by the method
presented in Ref.\cite{miyaguchi13}.

\section {Dimensionless forms}\label{app:nondim}

For numerical integration, the Markovian equations of motion
[Eqs.~(\ref{e.gle.wo.fd.markov.1}) and (\ref{e.gle.w.fd.markov.1})] are
discretized as
\begin{align}
  \label{e.app.dr/dt}
  d \bm{r}
  &=
  -\beta D\bm{F}_{e}dt -\beta D\sum_{i=0}^{N-1} k_i(\bm{r}-\bm{r}_i) dt
  + \sqrt{2D dt}\bm{\xi}(t),\\[0.1cm]
  \label{e.app.dr_i/dt}
  d \bm{r}_i 
  &=
  - \beta D_i(t)k_i (\bm{r}_i- \bm{r})dt + \sqrt{2D_i(t)dt} \bm{\eta}_i(t). 
\end{align}
Hereafter, we assume that $D_i(t)$ [or equivalently $\nu_i(t)$] is a two-state
process given by Eq.~(\ref{e.glefd.two-state}) and that the sojourn-time
distribution of the fast state $\rho^+(\tau)$ is given by a distribution with
mean $\mu_+$.

Then, Eqs.~(\ref{e.app.dr/dt}) and (\ref{e.app.dr_i/dt}) can be made
dimensionless by using transformations
\begin{align}
  \label{e.app.nondimensionalize}
  \tilde{\bm{r}}(\tilde{t}) &= \frac {\bm{r}(t)}{\sqrt{D\mu_+}}, \quad
  \tilde{\bm{r}}_i(\tilde{t}) = \frac {\bm{r}_i(t)}{\sqrt{D\mu_+}},
  \\[0.1cm]
  \tilde{t} &= \frac {t}{\mu_+}, \quad
  \tilde{k}_i = \beta D \mu_+ k_i,
  \\[0.1cm]
  \tilde{\bm{F}}_e &= \beta \sqrt{D \mu_+}\bm{F}_e, \quad
  \tilde{D}_i(\tilde{t}) = \frac {D_i(t)}{D}.
\end{align}
In the figures of this article, theoretical and numerical results are presented
with these units. The dimensionless equations of motion are then given by
\begin{align}
  \label{e.app.tilde.dr/dt}
  d \tilde{\bm{r}}
  &=
  -\tilde{\bm{F}}_{e}d\tilde{t}
  - \sum_{i=0}^{N-1} \tilde{k}_i(\tilde{\bm{r}}-\tilde{\bm{r}}_i) d\tilde{t}
  + \sqrt{2d\tilde{t}}\bm{\xi}(\tilde{t}),\\[0.1cm]
  \label{e.app.tilde.dr_i/dt}
  d \tilde{\bm{r}}_i 
  &=
  - \tilde{D}_i(\tilde{t})\tilde{k}_i (\tilde{\bm{r}}_i- \tilde{\bm{r}})dt
  + \sqrt{2\tilde{D}_i(\tilde{t})d\tilde{t}} \bm{\eta}_i(\tilde{t}). 
\end{align}
The sojourn time $\tau$, and its distributions $\rho^{\pm}(\tau)$ and
$\rho^{\pm, \mathrm{eq}}(\tau)$ for the two-state diffusivity models are made
dimensionless by transforms
\begin{equation}
  \tilde{\tau} = \frac {\tau}{\mu_+}, \quad
  \tilde{\rho}^{\pm}(\tilde{\tau}) = \mu_+ \rho^{\pm} (\tau), \quad
  \tilde{\rho}^{\pm, \mathrm{eq}}(\tilde{\tau})
  = \mu_+ \rho^{\pm, \mathrm{eq}} (\tau).
\end{equation}

As a numerical scheme to integrate Eqs.~(\ref{e.app.tilde.dr/dt}) and
(\ref{e.app.tilde.dr_i/dt}), the Euler method is employed \cite{kloeden11}.  In
addition, all the numerical simulations presented in this article are carried
out for one-dimensional systems $n=1$.

%section {bibliography}

%\bibliography{paper}


\end {document}